\documentclass[reprint,showpacs,superscriptaddress,english,twocolumn,notitlepage,longbibliography,nofootinbib]{revtex4-1}
\usepackage[utf8]{inputenc}
\usepackage{color}
\usepackage{array}
\usepackage{verbatim}
\usepackage{multirow}
\usepackage{amsmath}
\usepackage{amssymb}
\usepackage{dsfont}
\usepackage{graphicx}
\usepackage{esint}
\usepackage[unicode=true,
 bookmarks=true,bookmarksnumbered=true,bookmarksopen=false,
 breaklinks=true,pdfborder={0 0 0},backref=false,colorlinks=true]
 {hyperref}
\usepackage{enumitem}
\usepackage{titlesec}
\usepackage{cleveref}
\usepackage{longtable}
\usepackage{bbm}
\usepackage[normalem]{ulem}
\usepackage{listings}

\definecolor{mygreen}{RGB}{28,172,0} % color values Red, Green, Blue
\definecolor{mylilas}{RGB}{170,55,241}

\hypersetup{
citecolor={blue},
urlcolor={magenta}
}

\newcommand{\PreserveBackslash}[1]{\let\temp=\\#1\let\\=\temp}
\newcolumntype{C}[1]{>{\PreserveBackslash\centering}p{#1}}
\newcolumntype{R}[1]{>{\PreserveBackslash\raggedleft}p{#1}}
\newcolumntype{L}[1]{>{\PreserveBackslash\raggedright}p{#1}}

\usepackage{cleveref}
\crefname{appendix}{Appendix}{Appendices}
\crefname{equation}{Eq.}{Eqs.}
\crefname{figure}{Fig.}{Figs.}
\crefname{table}{Table}{Tables}
\crefname{section}{Section}{Sections}
\renewcommand{\paragraph}[1]{\vspace{0.2cm}{\bf \textit{#1}}}
\def\ie{{\it i.e.},\ }
\def\eg{{\it e.g.},\ }

\newcommand{\mbf}{\mathbf}
\newcommand{\mbb}{\mathbb}
\newcommand{\mcl}{\mathcal}

\newcommand{\mrm}{\mathrm}

\def\pare#1{\left( #1 \right)}

\def\bra#1{\langle #1 |}
\def\ket#1{| #1 \rangle}

\def\inn#1{\langle #1 \rangle}

\def\nono{\nonumber}

\def\pt{\partial}

\def\sgn{\mathrm{sgn}}

\def\kk{\mathbf{k}}
\def\ee{\epsilon}
\def\rr{\mathbf{r}}
\def\qq{\mathbf{q}}
\def\aa{\mathbf{a}}
\def\bb{\mathbf{b}}
\def\GG{\mathbf{G}}

\def\QQ{\mathbf{Q}}
\def\RR{\mathbf{R}}
\def\tt{\mathbf{t}}

\def\PH{\mathcal{P}}
\def\bsigma{\boldsymbol{\sigma}}
\def\ee{\epsilon}
\def\cH{\mathcal{H}}
\def\TRS{T}

\def\cA{\boldsymbol{\mathcal{A}}}
\def\cF{\mathcal{F}}

\def\MR{moir\'e}

\begin{document}
\title{Twisted Bilayer Graphene II: Stable Symmetry Anomaly in Twisted Bilayer Graphene}
\author{Zhi-Da Song}
\affiliation{Department of Physics, Princeton University, Princeton, New Jersey 08544, USA}
\author{Biao Lian}
\affiliation{Department of Physics, Princeton University, Princeton, New Jersey 08544, USA}
\author{Nicolas Regnault}
\affiliation{Department of Physics, Princeton University, Princeton, New Jersey 08544, USA}
\affiliation{Laboratoire de Physique de l'Ecole normale superieure, ENS, Universit\'e PSL, CNRS,
Sorbonne Universit\'e, Universit\'e Paris-Diderot, Sorbonne Paris Cit\'e, Paris, France}
\author{B. Andrei Bernevig}
\email{bernevig@princeton.edu}
\affiliation{Department of Physics, Princeton University, Princeton, New Jersey 08544, USA}

\date{\today}

\begin{abstract}
We show that the entire continuous model of twisted bilayer graphene (TBG) (and not just the two active bands) with particle-hole symmetry is anomalous and hence incompatible with lattice models. Previous works, 
\eg [\href{https://link.aps.org/doi/10.1103/PhysRevLett.123.036401}{Song et al., Phys. Rev. Lett. 123, 036401 (2019)}],  [\href{https://link.aps.org/doi/10.1103/PhysRevX.9.021013}{Ahn et al., Phys. Rev. X 9, 021013 (2019)}], [\href{https://link.aps.org/doi/10.1103/PhysRevB.99.195455}{Po et al., Phys. Rev. B 99, 195455 (2019)}], and others [\href{https://journals.aps.org/prx/abstract/10.1103/PhysRevX.8.031088}{Kang et al. Phys. Rev. X 8, 031088 (2018)}, \href{https://journals.aps.org/prx/abstract/10.1103/PhysRevX.8.031087}{Koshino et al., Phys. Rev. X 8, 031087 (2018)}, \href{https://journals.aps.org/prb/abstract/10.1103/PhysRevB.99.155415}{Liu et al., Phys. Rev. B 99, 155415 (2019)}, \href{https://journals.aps.org/prb/abstract/10.1103/PhysRevB.98.085435}{Zou et al., Phys. Rev. B 98, 085435 (2018)}]
%\cite{kang_symmetry_2018,koshino_maximally_2018,liu2019pseudo,zou2018}
found that the two flat bands in TBG possess a fragile topology protected by the $C_{2z}\TRS$ symmetry.
[\href{https://link.aps.org/doi/10.1103/PhysRevLett.123.036401}{Song et al., Phys. Rev. Lett. 123, 036401 (2019)}] also pointed out an approximate particle-hole symmetry ($\PH$) in the continuous model of TBG. 
In this work, we numerically confirm that $\PH$ is indeed a good approximation for TBG and show that the fragile topology of the two flat bands is enhanced to a $\PH$-protected stable topology. This stable topology implies $4l+2$ ($l\in\mbb{N}$) Dirac points between the middle two bands.
The $\PH$-protected stable topology is robust against arbitrary gap closings between the middle two bands the other bands.
We further show that, remarkably, this $\PH$-protected stable topology, as well as the corresponding $4l+2$ Dirac points, cannot be realized in lattice models that preserve both $C_{2z}\TRS$ and $\PH$ symmetries. In other words, the continuous model of TBG is anomalous and cannot be realized on lattices.
Two other topology related topics, with consequences for the interacting TBG problem, \ie the choice of Chern band basis in the two flat bands and the perfect metal phase of TBG in the so-called second chiral limit, are also discussed.
\end{abstract}

\maketitle

\section{Introduction}

TBG at the first magic angle ($\theta\approx 1.05^\circ$) exhibits a group of two almost exactly flat bands \cite{bistritzer_moire_2011}.
Due to the interesting interaction insulating and conducting states \cite{cao_correlated_2018,Efimkin2018TBG,xie2019spectroscopic,das2020symmetry,po_origin_2018,dodaro2018phases,yuan2018model,ochi_possible_2018,xux2018,venderbos2018,kang_strong_2019,liu2019quantum,jiang_charge_2019,choi_imaging_2019,polshyn_linear_2019,pixley2019,Xie2020TBG,bultinck_ground_2020,nuckolls_chern_2020,wu_chern_2020,saito2020,wong_cascade_2020,zondiner_cascade_2020,sharpe_emergent_2019,serlin_QAH_2019,bultinck2020,saito2020isospin,kang_nonabelian_2020,soejima2020efficient,cao_strange_2020,kwan2020orbital}, 
superconductor states \cite{cao_unconventional_2018,lu2019superconductors,yankowitz2019tuning,Wu2018TBG-BCS,xu2018topological,liu2018chiral,isobe2018unconventional,guinea2018,gonzalez2019kohn,Lian2019TBG,you2019,xie_superfluid_2020,saito_independent_2020,stepanov_interplay_2020,arora_2020,khalaf_charged_2020,wu_collective_2020,julku_superfluid_2020,knig2020spin}, 
and single-particle topology \cite{kang_symmetry_2018,koshino_maximally_2018,ahn_failure_2019,po_faithful_2019,song_all_2019,liu2019pseudo,tarnopolsky_origin_2019,fu2018magicangle,zhang2019nearly,bultinck_ground_2020,lian2020,lu2020fingerprints,padhi2020transport,Jonah_hofstadter_2020,Wilson2020TBG} in the flat bands, TBG represents one of the most versatile physical systems of recent years 
\cite{bistritzer_moire_2011,cao_correlated_2018,cao_unconventional_2018, lu2019superconductors, yankowitz2019tuning, sharpe_emergent_2019, saito_independent_2020, stepanov_interplay_2020, liu2020tuning, arora_2020, serlin_QAH_2019, cao_strange_2020, polshyn_linear_2019,  xie2019spectroscopic, choi_imaging_2019, kerelsky_2019_stm, jiang_charge_2019,  wong_cascade_2020, zondiner_cascade_2020,  nuckolls_chern_2020, choi2020tracing, saito2020,das2020symmetry, wu_chern_2020,park2020flavour, saito2020isospin,rozen2020entropic, lu2020fingerprints, burg_correlated_2019,shen_correlated_2020, cao_tunable_2020, liu_spin-polarized_2019, chen_evidence_2019, chen_signatures_2019, chen_tunable_2020, burg2020evidence, tarnopolsky_origin_2019, zou2018, fu2018magicangle, liu2019pseudo, Efimkin2018TBG, kang_symmetry_2018, song_all_2019,po_faithful_2019,ahn_failure_2019,Slager2019WL, hejazi_multiple_2019, lian2020, hejazi_landau_2019, padhi2020transport, xu2018topological,  koshino_maximally_2018, ochi_possible_2018, xux2018, guinea2018, venderbos2018, you2019,  wu_collective_2020, Lian2019TBG,Wu2018TBG-BCS, isobe2018unconventional,liu2018chiral, bultinck2020, zhang2019nearly, liu2019quantum,  wux2018b, thomson2018triangular,  dodaro2018phases, gonzalez2019kohn, yuan2018model,kang_strong_2019,bultinck_ground_2020,seo_ferro_2019, hejazi2020hybrid, khalaf_charged_2020,po_origin_2018,xie_superfluid_2020,julku_superfluid_2020, hu2019_superfluid, kang_nonabelian_2020, soejima2020efficient, pixley2019, knig2020spin, christos2020superconductivity,lewandowski2020pairing, xie_HF_2020,liu2020theories, cea_band_2020,zhang_HF_2020,liu2020nematic, daliao_VBO_2019,daliao2020correlation, classen2019competing, kennes2018strong, eugenio2020dmrg, huang2020deconstructing, huang2019antiferromagnetically,guo2018pairing, ledwith2020, repellin_EDDMRG_2020,abouelkomsan2020,repellin_FCI_2020, vafek2020hidden,fernandes_nematic_2020, Wilson2020TBG, wang2020chiral, ourpaper1,ourpaper3,ourpaper4,ourpaper5,ourpaper6}. 
Refs. \cite{ahn_failure_2019, song_all_2019} showed that the $C_{2z}\TRS$ symmetry of TBG protects a fragile topology \cite{po_fragile_2018,cano_fragile_2018,Slager2019WL,Else2019fragile,Juan2020fragile,Aris2020fragile} of the two flat bands, which is characterized by a $\mathbb{Z}$-valued winding number. 
The fragile topology manifests itself as a topological obstruction for exponentially decaying Wannier functions satisfying $C_{2z}\TRS$ symmetry for the two flat bands.
However, the Wannier obstruction can be removed by adding trivial bands into the consideration \cite{po_fragile_2018,cano_fragile_2018,Slager2019WL}. 
For example, Ref. \cite{po_faithful_2019} showed explicitly that symmetric Wannier functions can be constructed if certain additional orbitals are coupled the fragile topological band protected by $C_{2z}\TRS$. However, the papers arguing for a trivialization of the bands \cite{po_faithful_2019,ahn_failure_2019} neglected one (approximate) symmetry of the TBG model \cite{bistritzer_moire_2011}. 

The Bistritzer MacDonald (BM) model \cite{bistritzer_moire_2011} of TBG has an approximate particle-hole symmetry $\PH$ first pointed out in Ref. \cite{song_all_2019}. It was already pointed out in Ref. \cite{song_all_2019} that with this approximate symmetry, there seems to be a further, stable topology in TBG, but this result was not further expanded.
We here numerically confirm that the error - on the wavefunctions - of the $\PH$ symmetry (defined in \cref{sec:symm}) in the BM model of TBG is extremely small ($<0.01$).
Thus we count $\PH$ symmetry as a good approximation for the low energy physics in TBG.
We prove that if the $C_{2z}\TRS$ protected winding number of the two flat bands is odd (true in TBG), then the two flat bands have a stable topology protected by $\PH$, which is characterized by a $\mbb{Z}_2$ invariant $\delta$.
In contrast to the fragile topological bands, which can be trivialized by being coupled to certain trivial bands, the $\mbb{Z}_2$ topology, as well as the Wannier obstruction implied by the $\mbb{Z}_2$ invariant, is stable against adding trivial bands that preserve the $\PH$ symmetry. 
We further proved that, in the presence of $C_{2z}\TRS$ and $\PH$, the $\mbb{Z}_2$ invariant $\delta$ of $2M$ particle-hole symmetric bands $\ee_{-M}(\kk)\cdots \ee_{-1}(\kk),\ee_{1}(\kk)\cdots \ee_{M}(\kk)$ is related to the number of Dirac points $N_D$ between $\ee_{-1}(\kk)$ and $\ee_{1}(\kk)$ in the first Brillouin zone (BZ) as $\delta = \frac{N_D}2$ mod 2, provided that the $2M$ bands are gapped from higher and lower bands.
Here $\ee_n(\kk)$ ($\ee_{-n}(\kk)$) is the $n$-th positive (negative) band. 
Therefore, as long as  $4l+2$ ($l\in\mbb{N}$) Dirac points exist between $\ee_{-1}(\kk)$ and $\ee_{1}(\kk)$ we find that $2M, \;\forall M \in \mbb{N}$ particle-hole symmetric bands (separate in energy from the $M+1 ,M+2 \ldots$ and $\ldots, -M-2, -M-1$ bands) have $\delta=1$ and hence are topologically nontrivial. 
The feature of TBG that arbitrary $2M$ bands are topological is inconsistent with lattice models with $C_{2z}\TRS$ and $\PH$ symmetry.
In a lattice model, if $2M$ is large enough, \eg equals to the number of orbitals in the model, the $2M$ bands have to be topologically trivial because they span the  Hilbert space of the local orbitals. 
Therefore, the $\mbb{Z}_2$ topology, and the $4l+2$ Dirac points accordingly, cannot be realized in lattice models with finite number of orbitals. 
We hence call the $\mbb{Z}_2$ topology an anomaly of the $C_{2z}\TRS$ and $\PH$ symmetries. We further note that this implies that the many-body U(4) and U(4) $\times $ U(4) symmetries \cite{ourpaper3,ourpaper4,ourpaper5} are incompatible with a lattice model and hence anomalous. It also implies that the lattice models build to model TBG \cite{po_faithful_2019,kang_symmetry_2018,koshino_maximally_2018,bultinck_ground_2020} have to break the $\PH$ symmetry or the $C_{2z}T$ symmetry of the TBG model.

This paper is organized as follows.
In \cref{sec:TBG}, we present a review of the BM model of TBG and summarize its symmetries.
The error of the approximate particle-hole symmetry $\PH$ is defined and is confirmed as being small ($<0.01$).
In \cref{sec:Z2top}, we prove that the $\PH$ symmetry protects a stable $\mbb{Z}_2$ topological state.
In \cref{sec:no-go}, a no-go theorem of the $\mbb{Z}_2$ topology is proved for lattice models with the $C_{2z}\TRS$ and $\PH$ symmetries. 
The relation between the $\mbb{Z}_2$ invariant and the number of Dirac points is also established in this section.
In \cref{sec:Chern}, we show that, when the two flat bands are gapped from the other bands, there is natural choice of Chern band basis (with opposite Chern numbers) in the two flat bands. 
The Chern band basis is used in our interacting works \cite{ourpaper3, ourpaper4, ourpaper5, ourpaper6} on TBG.
In \cref{sec:S2}, we show that, in the so-called second chiral limit, defined in \cite{ourpaper3} as the second limit having an interacting extended U(4) $\times$ U(4) symmetry, the symmetry anomaly of TBG manifests as a perfect metal phase, where all the bands are connected to each other. 
A brief summary of this work is given in \cref{sec:conclusion}.

\section{The BM model of twisted bilayer graphene and its symmetries} \label{sec:TBG}

We first present a short  review of the BM model. A more detailed account can be found in supplementary material of Ref. \cite{song_all_2019}.

\subsection{A brief review of the BM model}\label{sec:TBG-review}
TBG is an engineered material of two graphene layers twisted by a small angle $\theta$ from each other. 
The band structure of each of the two layers exhibits two Dirac points at the $K$ and $K'$ momenta in the single layer Brillouin zone (BZ), respectively; the two Dirac points are related by time-reversal $\TRS$.
Thus the band structure of TBG exhibits four Dirac points: two from the top layer and the other two from the bottom layer.
When $\theta$ is small such that the interlayer coupling is smooth in real space - with a length scale much larger than the atom distances - the graphene valley ($K$ and $K'$) is a good quantum number of low energy states of TBG \cite{bistritzer_moire_2011}.
In this case, the states around $K$ ($K'$) in the top layer only couple to the states around $K$ ($K'$) in the bottom layer. 
Therefore, the low energy band structure of TBG decomposes into two independent graphene valleys, and each valley has two Dirac points originated from the two layers, respectively.
In this work, we will focus on the valley $K$.
The bands in the other valley $K'$ can be obtained by acting $T$ on the bands in the valley $K$.

We assume the top single graphene layer is rotated from the $x$-direction by an angle $\frac{\theta}2$ (rotation axis is $z$). 
Thus the Dirac Hamiltonian around $K$ in the top layer is $-i v_F\pt_x (\cos\frac{\theta}2\sigma_x-\sin\frac{\theta}2\sigma_y) -i v_F \pt_y (\cos\frac{\theta}2\sigma_y+\sin\frac{\theta}2\sigma_x) \approx -iv_F \pt_\rr\cdot\bsigma + i\frac{\theta}2 v_F \pt_\rr\times \bsigma$, where $v_F$ is the Fermi-velocity of single-layer graphene and $\bsigma=(\sigma_x,\sigma_y)$ are Pauli matrices representing the A/B sublattices of graphene.
The bottom layer is rotated from the $x$-direction by an angle $-\frac{\theta}2$.
Correspondingly, the Dirac Hamiltonian around $K$ in the bottom layer is $ -iv_F \pt_\rr\cdot\bsigma  -i\theta v_F \pt_\rr\times \bsigma$.
The interlayer coupling is encoded in a position dependent matrix $T(\rr)$, where $\rr=(x,y)$, such that the Hamiltonian of TBG, to linear order of $\theta$, can be written as 
\begin{equation}
H(\rr) = 
-iv_F \pare{\tau_0 \pt_\rr\cdot\bsigma - \frac{\theta}2 \tau_z \pt_\rr\times \bsigma} + 
\begin{pmatrix}
 0  &  T(\rr)  \\
 T^\dagger(\rr) &  0 
\end{pmatrix} . \label{eq:Ham}
\end{equation}
Here $\tau_0$ and $\tau_z$ are the two-by-two identity matrix and the third Pauli matrix for the layer degree of freedom, respectively.
According to Ref. \cite{bistritzer_moire_2011}, when $\theta$ is small ($\sim 1^\circ$), $T(\rr)$ forms a smooth \MR  potential:
\begin{equation}
    T(\rr) = \sum_{i=1}^3 e^{-i\qq_i \cdot \rr} T_i, \label{eq:Tr}
\end{equation}
where $\qq_i$'s are $\qq_1=k_D(0,-1)$, $\qq_2=k_D(\frac{\sqrt3}2,\frac12)$, $\qq_3=k_D(-\frac{\sqrt3}2,\frac12)$, with $k_D=2|K|\sin\frac{\theta}2$ being the distance between $K$ momenta in the two layers, and $T_i$'s are 
\begin{equation} \label{eq:Ti-def}
\begin{aligned}
    T_i =& w_0 \sigma_0 + w_1\Big[\sigma_x\cos\frac{2\pi(i-1)}3+ \sigma_y\sin\frac{2\pi(i-1)}3\Big],
\end{aligned}
\end{equation}
where $w_0$ and $w_1$ are two constant parameters.
Since the $w_0$ term contributes to the diagonal elements, it represents the interlayer coupling between the A(B) sublattice of the top layer and the A(B) sublattice of the bottom layer. 
Similarly, the $w_1$ term only contributes to the off-diagonal elements, it is thus associated to the interlayer coupling between A(B) sublattice of the top layer and B(A) sublattice of the bottom layer. 
%Due to the corrugation effect, as explained in next paragraph, there is $w_0<w_1$ \cite{koshino_maximally_2018}. 

\begin{figure*}[t]
\centering
\includegraphics[width=1\linewidth]{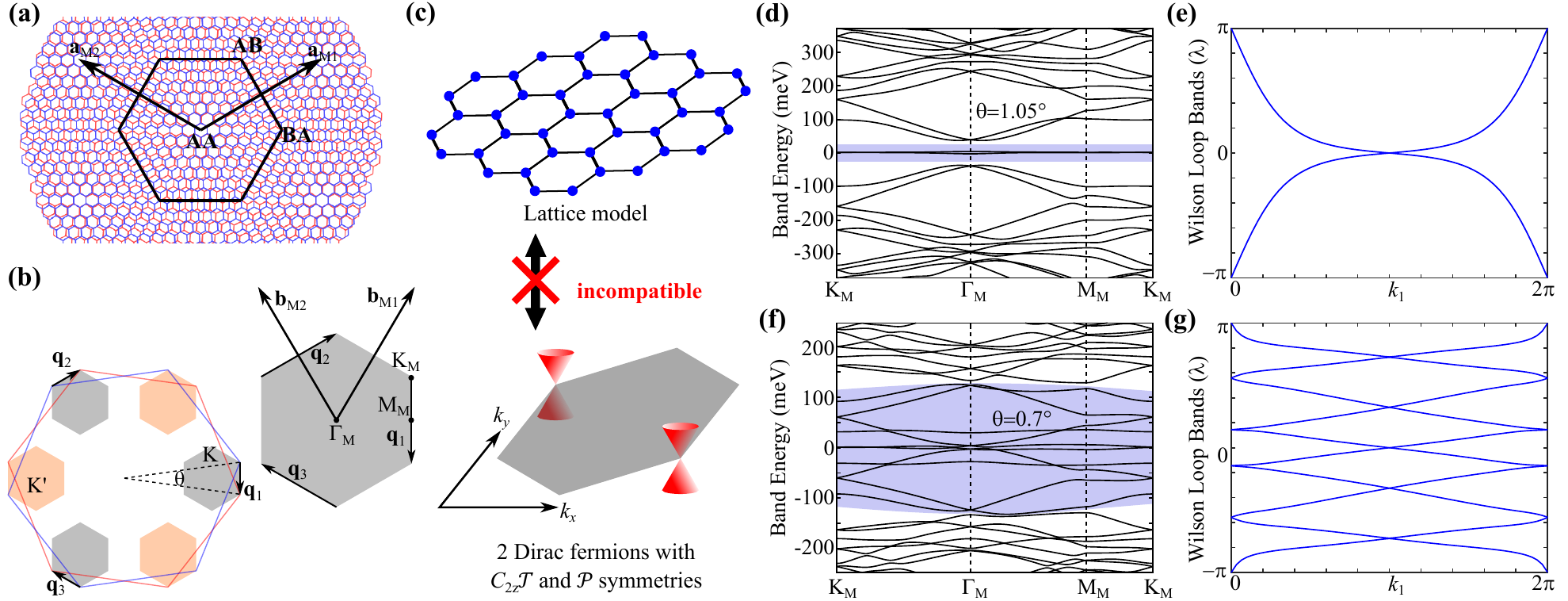}    
\caption{The lattice, symmetry anomaly, band structures, and Wilson loop bands of TBG. 
\textbf{(a)} The \MR unit cell, where the blue sheet and the red sheet represent the top and bottom layers, respectively. In the AA, AB, BA regions, the A sublattice of the top layer are located above the A sublattice, the B sublattice, and the hexagon center of the bottom layer, respectively. 
\textbf{(b)} The \MR Brillouin zone. Left: The grey and yellow hexagons represent the \MR Brillouin zone for the graphene valleys $K$ and $K'$, respectively. Right: The reciprocal lattices and the high symmetry momenta of the \MR Brillouin zone in graphene valley $K$.
\textbf{(c)} $4l+2$ ($l\in\mbb{N}$) Dirac points cannot be realized in lattice models with $C_{2z}\TRS$ and $\PH$ symmetries. 
\textbf{(d-e)} The band structure and Wilson loop bands of the middle two bands (shaded) at $\theta=1.05^\circ$. The crossings in Wilson loop bands are protected by $C_{2z}\TRS$ and/or the approximate $\PH$. Each Wilson loop operator is integrated along $\bb_{M2}$ and the spectrum is plotted along $\bb_{M1}$.
\textbf{(f-g)} The band structure and Wilson loop bands of the middle ten bands (shaded) at $\theta=0.7^\circ$. The crossings at $\lambda=0,\pi$ in the Wilson loop bands are protected by $C_{2z}\TRS$ and/or by the approximate $\PH$; the double degeneracies with $\lambda\neq 0,\pi$ at $k_1=0,\pi$ are protected by the approximate $\PH$. These double-degeneracies guarantee Wilson loop flow for any bands with $4n+2$ Dirac nodes at zero energy.
In fact, we have kept the $\PH$-breaking term $i\theta v_F \tau_z \pt_\rr \times\bsigma$ (\cref{eq:Ham}) in the calculations used to generate this plot, which would split the double degeneracies in principle.
However, the splittings are almost invisible by eye in the plot, implying that the $\PH$ symmetry is a good approximation. The degeneracies are exact when $P$ is exact.
The parameters of Hamiltonian used in (d-g) are $v_F=5.944{\rm eV \cdot \mathring{ A} }$, $|K|=1.703\mathring{\rm A}^{-1}$, $w_1=110{\rm meV}$, $w_0=0.7w_1$.
}
\label{fig:TBG}
\end{figure*}

The \MR potential (\cref{eq:Tr}) is invariant (up to a gauge transformation) under the translations
$\aa_{M1} = \frac{2\pi}{k_D} (\frac1{\sqrt3}, \frac13)$,
$\aa_{M2} = \frac{2\pi}{k_D} (-\frac1{\sqrt3}, \frac13)$.
The translation symmetry of the \MR potential is manifest in real space (\cref{fig:TBG}a).
The corresponding reciprocal lattice bases are $\bb_{M1}=\qq_2-\qq_1$, $\bb_{M2}=\qq_3-\qq_1$ (\cref{fig:TBG}b).
The unit cell spanned by $\aa_{M1}$ and $\aa_{M2}$ is referred to as the \MR unit cell. 
Each \MR unit cell has one AA region, one AB region, and one BA region. 
In the AA region, the  A(B)  sublattice of the top layer sit on top of the A(B) sublattice of the bottom layer; in the AB region, the A and B sublattices of the top layer sit on top of the B sublattices and the empty hexagon centers of the bottom layer, respectively; in the BA region, the B and A sublattices of the upper layer sit on top of the A sublattices and the empty hexagon centers of the lower layer, respectively.
First principle calculations show that the two layers are corrugated in the $z$-direction \cite{Uchida_corrugation,Wijk_corrugation,dai_corrugation,jain_corrugation}.
The distance between the two layers in the AA region is larger than the distance in the AB and BA regions.
Since $w_0$ and $w_1$ are mainly dominated by the couplings in the AA and AB/BA regions \cite{bistritzer_moire_2011}, respectively, this implies that, in the realistic model, $w_0$ is smaller than $w_1$ \cite{koshino_maximally_2018}. 
In Figs.~\ref{fig:TBG}d and~\ref{fig:TBG}f, we show the the band structures for two different twist angles $\theta=1.05^\circ,0.7^\circ$.
The parameters are set as $v_F=5.944{\rm eV \cdot \mathring{ A} }$, $|K|=1.703\mathring{\rm A}^{-1}$, $w_1=110{\rm meV}$, $w_0=0.7w_1$.

\subsection{Symmetries of the BM model} \label{sec:symm}

The model \cref{eq:Ham} has several point group symmetries: (i) $C_{2z}\TRS=\sigma_x K$, where $K$ is the complex conjugation,
% \begin{equation}
% C_{2z}\TRS H(\rr) (C_{2z}\TRS)^{-1} = H(-\rr),
% \end{equation}
(ii) $C_{3z}=e^{i\frac{2\pi}3\sigma_z}$,
% \begin{equation}
% C_{3z} H(\rr) C_{3z}^{\dagger} = H(C_{3z}\rr),
% \end{equation}
(iii) $C_{2x}=\tau_x\sigma_x$.
One can verify that the Hamiltonian is invariant under these symmetries. 
Notice that the single-graphene-valley Hamiltonian does not have the $C_{2z}$ rotation and the time-reversal $\TRS$ symmetries since they map one graphene valley to the other.
The three crystalline symmetries and the \MR translations generate the magnetic space group $P6'2'2$ (\#177.151 in the BNS setting \cite{Bilbao-MSG}) \cite{song_all_2019}. 

We define a unitary particle-hole operation $P=i\tau_y$, which transforms the position as $\rr\to -\rr$  \cite{song_all_2019}.
Here $\tau_{x,y,z}$ are Pauli matrices representing the layer degree of freedom.
Under $P$ the Hamiltonian transforms as
\begin{equation}
P H (\rr) P^\dagger = - H(-\rr) + i \theta v_F \tau_z  \pt_\rr\times \bsigma . \label{eq:P-action}
\end{equation}
The second term on the right hand side is in linear order of $\theta$.
It approaches zero when $\theta\to 0$. 
While the other terms in $H(\rr)$ do not vanish in the $\theta\to 0$ limit.
Thus it is safe to ignore this term in the small angle limit.
To be specific, when $\theta\sim 1^\circ$, this term is of order $0.018 v_F k_D$ and hence is much smaller than the energy scale of the low energy physics, which is of order $v_Fk_D$.
Therefore, $P$ is an emergent anticommuting symmetry when $\theta$ is small.
It satisfies the algebra \cite{song_all_2019}:
{
\begin{equation}
[C_{2z}\TRS,P]=
[C_{3z},P]=0,\;
\{C_{2x},P\}=0,\;
P^2=-1. \label{eq:P-algebra}
\end{equation}
}%
For later convenience, we define an anti-unitary particle operation $\PH=P C_{2z}\TRS = i\tau_y \sigma_x K$, which is local in real space and satisfies $\PH^2=-1$.
It acts on the Hamiltonian as 
\begin{equation}
\PH H (\rr) \PH^{-1} = - H(\rr) + i \theta v_F \tau_z  \pt_\rr\times \bsigma .
\end{equation}
As discussed in details in \cref{app:PHbreaking}, the $\kk$-dependence in the interlayer coupling may also cause a breaking of the emergent $\PH$ symmetry. 
We can also define a chiral operation $C=\sigma_z$, which is local in real space. \cite{tarnopolsky_origin_2019}.
Under $C$ the Hamiltonian transforms as 
\begin{equation}
C H(\rr) C^\dagger = -H(\rr) + 2 w_0 \tau_x \sigma_0 \sum_{i=1}^3 e^{-i\qq_i\cdot\rr}. \label{eq:chiral}
\end{equation}
In the so-called chiral limit \cite{tarnopolsky_origin_2019}, \ie $w_0=0$, the second term on the right hand of side vanishes and hence $C$ become an emergent anticommuting symmetry.
The chiral symmetry satisfies the algebra
{
\begin{equation}
\{C_{2z}\TRS,C\}=
\{C_{2x},C\}=0,\;
[C_{3z},C]=
[P,C]=0,\;
C^2=1. \label{eq:S-algebra}
\end{equation}
}%

\begin{figure}[t]
\centering
\includegraphics[width=\linewidth]{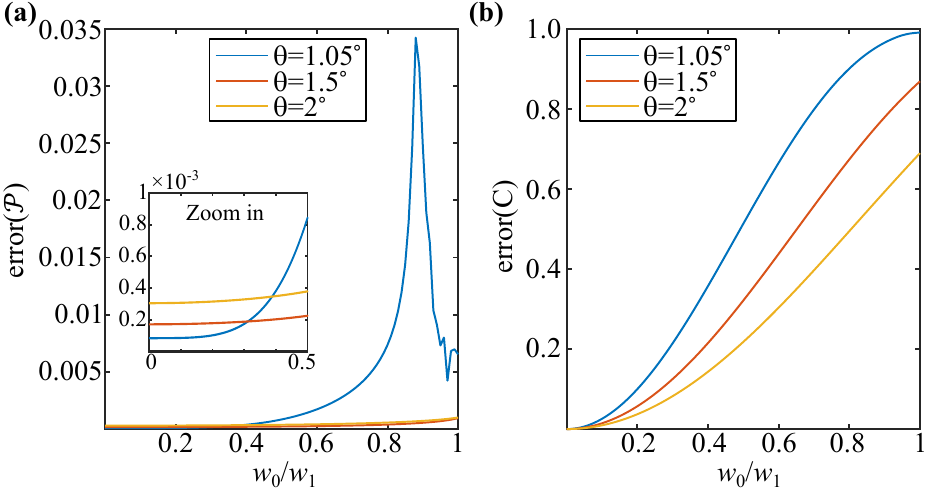}
\caption{Errors of the approximate symmetries $\PH$ (a) and $C$ (b) on the wavefunctions (as defined in Eqs.~\ref{eq:errorPH} and~\ref{eq:errorC}) in TBG as functions of $w_0/w_1$. Here we change $w_0$ while keeping $w_1$ fixed (110meV). The errors are shown for different values of the twist angle $\theta= 1.05^\circ,1.5^\circ, 2^\circ$.}
\label{fig:error}
\end{figure}

We numerically checked how much $\PH$ and $C$ are broken in the \emph{wavefunctions} of the model \cref{eq:Ham}.
To be specific, we define the errors of the two symmetries in the two flat bands as
\begin{equation}
{\rm error}(\PH) = {1-  \frac{1}{\Omega_M} \int d^2\kk |\inn{u_{1,-\kk}|\PH|u_{-1,\kk}}|^2},\label{eq:errorPH}
\end{equation}
\begin{equation}
{\rm error}(C) = {1- \frac{1}{\Omega_M} \int d^2\kk |\inn{u_{1,\kk}|C|u_{-1,\kk}}|^2},\label{eq:errorC}
\end{equation}
respectively, where $\ket{u_{-1,\kk}}$ and $\ket{u_{1,\kk}}$ are the periodic parts of the Bloch states of the highest occupied band and the lowest empty band at charge neutrality, respectively, and $\Omega_M=|\bb_{M1}\times\bb_{M2}|=\frac{3\sqrt3}{2}k_D^2$ is the area of the Moire Brillouin zone.
When the two symmetries are exact, we have $|\inn{u_{1,-\kk}|\PH|u_{-1,\kk}}|=|\inn{u_{1,\kk}|C|u_{-1,\kk}}|=1$ and hence the errors are zero.
%Using the parameters introduced in the end of \cref{sec:TBG-review}, where $w_0=0.7w_1$, we find $\mrm{error}(\PH)\approx 0.0034$, $\mrm{error}(C)\approx 0.80$ at $\theta=1.05^\circ$.
%If we set $w_0=0.1w_1$ and keep all the other parameters unchanged, then the errors become $\mrm{error}(\PH)\approx 8.5\times10^{-5}$, $\mrm{error}(C)\approx 0.026$. 
Using the parameters $v_F=5.933\mrm{eV\cdot\mathring{A}}$, $|K|=1.703\mathring{A}^{-1}$, $w_1=110\mrm{meV}$, we plot $\mrm{error}(\PH)$ and $\mrm{error}(C)$ as functions of $w_0/w_1$ (with fixed $w_1$) for a few twist angles in \cref{fig:error}. 
For $\theta=1.05^\circ$, $\mrm{error}(\PH)$ is small ($<0.01$) for $w_0\le 0.82 w_1$, thus the $\PH$ symmetry is a good approximation for TBG, while the $C$ symmetry only starts being good (with error$<0.01$) for $w_0\le0.07w_1$.

\section{Stable topology protected by particle-hole  symmetry \texorpdfstring{$\PH$}{P}} \label{sec:Z2top}

\subsection{The Wilson loop \texorpdfstring{$\mathbb{Z}_2$}{Z2} invariant protected by \texorpdfstring{$\PH$}{P}}

We denote the Hamiltonian in momentum space as $H(\kk)$.
We assume the emergent anti-unitary particle-hole symmetry, \ie $\PH H(\kk) \PH^{-1} = -H(-\kk)$, and $\PH^2=-1$. As detailed in \cref{sec:symm}, $\PH =P C_{2z}\TRS$ is anti-unitary and squares to -1, and is the product of the unitary $P$ of Ref. \cite{song_all_2019} and $C_{2z}\TRS$. We denote the energy and the periodic part of Bloch state of the $n$-th band above (below) the zero energy as $\ee_{n}(\kk)$ ($\ee_{-n}(\kk)$) and $\ket{u_{n}(\kk)}$ ($\ket{u_{-n}(\kk)}$),  respectively.
As explained in \cref{app:Hk} and in Ref. \cite{song_all_2019}, $\ket{u_{n}(\kk)}$ satisfies the periodicity $\ket{u_n(\kk+\GG)}=V^\GG \ket{u_n(\kk)}$, with $\GG$ being a reciprocal lattice and $V^\GG$ a unitary matrix referred to as the embedding matrix.
Since $\PH$ anti-commutes with the Hamiltonian and flips the momentum, we have  $\ee_n(\kk) = - \ee_{-n}(-\kk)$.
The state $\PH \ket{u_n(\kk)}$ must have the momentum $-\kk$ and the energy $\ee_{-n}(-\kk)$.
In general, $\PH \ket{u_n(\kk)}$ is spanned by Bloch states at $-\kk$ as
\begin{equation}
    \PH \ket{u_n(\kk)} = \sum_{n'} \ket{u_{n'} (-\kk)} B_{n'n}^{(\PH)}(\kk), \label{eq:PH-sewing}
\end{equation}
where the summation over $n'$ is limited to those satisfying $\ee_{n'}(\kk)=-\ee_{n}(-\kk)$, and $B_{n'n}^{(\PH)}(\kk)$ is a unitary matrix referred to as the sewing matrix of $\PH$. 
$B^{(\PH)}(\kk)$ is periodic in momentum space, \ie $B^{(\PH)}(\kk+\GG)=B^{(\PH)}(\kk)$ \cite{alexandradinata2016topological,wang2016hourglass}.
Since $\PH^2=-1$, it should satisfy
\begin{equation}
B^{(\PH)}(-\kk) B^{(\PH)*}(\kk) = -1. \label{eq:PH-sewing-cond}
\end{equation}
Multiplying  $B^{(\PH)T}(\kk)$ on the right hand side of the above equation, we obtain
\begin{equation}
B^{(\PH)}(-\kk) = -B^{(\PH)T}(\kk). \label{eq:PH-sewing-cond2}
\end{equation}

We now prove that the $\PH$ symmetry protects a $\mathbb{Z}_2$ invariant for $2M$ particle-hole symmetric separate bands, \ie bands $\ee_{-M}(\kk), \ee_{-M+1}(\kk) \cdots \ee_{M}(\kk)$, gapped from higher and lower bands.
This proof is not limited to TBG but applies to any system having our anti-unitary $\PH$ symmetry. 
We introduce the matrix $U(\kk)=(\ket{u_{-M}(\kk)}, \ket{u_{-M+1}(\kk)} \cdots \ket{u_{M}(\kk)})$.
We parameterize $\kk$ as $k_1\bb_1+k_2\bb_2$, where $\bb_{1}$ and $\bb_2$ are the reciprocal lattice basis vectors.
Then we define the Wilson loop operator of the $2M$ bands for a given $k_1$ as
\begin{equation}
W(k_1) = \lim_{N\to \infty} \prod_{j=0}^{N-1} U^\dagger({k_1,j\frac{2\pi}N}) U({k_1,(j+1)\frac{2\pi}N}). \label{eq:WL}
\end{equation}
The order of the matrices in the product is given by $j$: matrices with larger $k_2$ ($=j\frac{2\pi}N$) always appear on the right hand side of matrices with smaller $k_2$.
Due to the periodicity of Bloch states, $W(k_1)$ is periodic..
Since $W(k_1)$ is unitary, its eigenvalues are phase factors $e^{i\lambda_n(k_x)}$ ($n=1\cdots 2M$), where $\lambda_n(k_1)$ ranges from $-\pi$ to $\pi$.
$\{\lambda_{n}(k_1)\}$ are called as the Wilson loop bands.
Topology is usually a result of Wilson loop flow, which in turn is a result of unavoidable crossings between Wilson loop bands.

\begin{figure}[t]
\centering
\includegraphics[width=\linewidth]{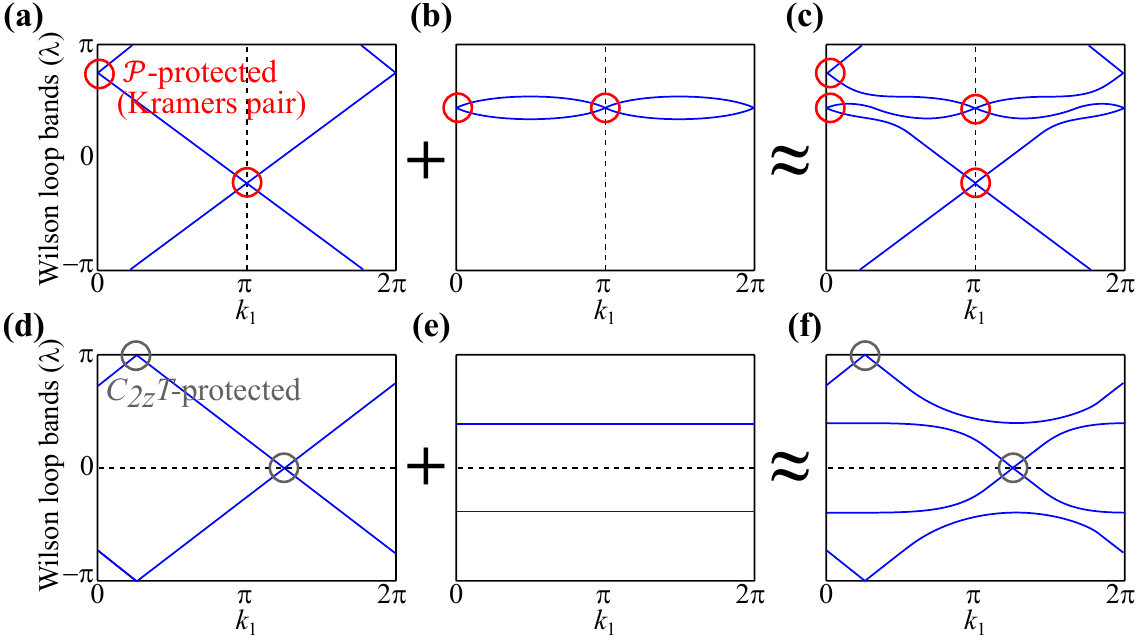}    
\caption{Comparison of Wilson loop windings protected by $\PH$ and $C_{2z}\TRS$.
\textbf{(a-c)} The Wilson loop bands with $\PH$. The crossings at $k_1=0,\pi$ are Kramers pairs protected by $\PH$. The $\mbb{Z}_2$ invariant (a) equals to 1 if the Wilson loop bands form a zigzag connection between $k_1=0$ and $k_1=\pi$  and (b) equals to 0 otherwise. The $\PH$-protected topology is stable against adding trivial bands: Coupling the nontrivial Wilson loop bands (a) to a trivial Wilson loop bands (b) yields a nontrivial Wilson loop bands (c).
\textbf{(d-f)} The Wilson loop bands with $C_{2z}\TRS$. The crossings at $\lambda=0,\pi$ are protected by $C_{2z}\TRS$. For a two-band system, the topology is nontrivial if the Wilson loop bands winds (d) and is trivial otherwise (e). The $C_{2z}\TRS$-protected topology is fragile: Coupling the nontrivial Wilson loop bands (d) to a trivial Wilson loop bands (f) yields a trivial Wilson loop bands (f). 
}
\label{fig:WL}
\end{figure}

We now prove that the Wilson loop bands are doubly degenerate at $k_1=0$ and $k_1=\pi$, as shown in \cref{fig:WL}a-c. In fact, we should heuristically expect this, since the Wilson loop respects $\PH$ as it contains all bands related by the particule-hole symmetry. Since $\PH$ is anti-unitary and squares to $-1$ is hence acts as spinful time-reversal, which we already know to enforce Kramers doublets in the Wilson loop spectrum \cite{yu_equivalent_2011,Aris2014WL}. Due to \cref{eq:PH-sewing}, we have 
\begin{align}
&\inn{ u_n(\kk) | u_{n'}(\kk') } = \inn{ \PH u_{n'}(\kk') | \PH u_{n}(\kk) }\nono\\
=& \sum_{m m'} B^{(\PH) *}_{m'n'}(\kk') \inn{ u_{m'}(-\kk') | u_{m}(-\kk) } B^{(\PH)}_{mn}(\kk). \label{eq:ukukp-P}
\end{align}
In the above equation we have made use of a property of anti-unitary symmetries: for any two states $\ket{\phi}$, $\ket{\psi}$ and an arbitrary anti-unitary operator $\mcl{O}$, we have $\inn{\phi|\psi} = \inn{\mcl{O} \psi|\mcl{O} \phi}$.
Substituting this relation into \cref{eq:WL} and using the periodicity relations $B^{(\PH)}(\kk+\GG)=B^{(\PH)}(\kk)$ and $\ket{u_n(\kk+\GG)}=V^\GG \ket{u_n(\kk)}$, we obtain
\begin{equation}
W(k_1) = B^{(\PH)T} (k_1,0) W^T(-k_1) B^{(\PH)*}(k_1,0). \label{eq:WL-sym-PH}
\end{equation}
Since $W(k_1)$ is periodic at $k_1$, $W(k_1)$ with $k_1=0,\pi$ are invariant under the particle-hole operation:
\begin{equation}
W(k_1) = B^{(\PH)T} (k_1,0) W^T(k_1) B^{(\PH)*}(k_1,0),\quad (k_1=0,\pi).
\end{equation}
It is this invariance that protects degeneracies of Wilson loop bands at $k_1=0,\pi$.
To see this, we parameterize the unitary matrix $W(k_1)$ as $e^{i\cH(k_1)}$ with $\cH(k_1)$ being a hermitian matrix periodic in $k_1$, called the Wilson Hamiltonian.
The eigenvalues of $\cH(k_1)$ form the Wilson loop bands.
We can define the particle-hole operator for $\cH(k_1)$ as $\tilde{\PH}(k_1)=B^{(\PH)}(k_1,0) K$ such that \cref{eq:WL-sym-PH} can be written as $\cH(k_1)=\tilde{\PH}(-k_1) \cH(-k_1) \tilde{\PH}^{-1}(-k_1)$.
We have $\tilde{\PH}(k_1)\tilde{\PH}(-k_1)=-1$ due to \cref{eq:PH-sewing-cond}.
It is worth noting that unlike the Hamiltonian $H(\kk)$ which anti-commutes with $\PH$, $\cH(k_1)$ commutes with $\tilde{\PH}$.
%Since $\cH(k_x)$ is periodic in $k_x$, $\cH(0)$ and $\cH(\pi)$ are invariant under $\tilde{\PH}$.
Because $\tilde{\PH}^2(k_1)=-1$ and $[\tilde{\PH},\cH(k_1)]=0$ for $k_1=0,\pi$, the Wilson loop bands - the eigenstates of the Wilson Hamiltonian - at $k_1=0,\pi$ form doublets due to the Kramers theorem.
% For example, an eigenvector $v$ of $\cH(0)$ is orthogonal to $\td{\PH} v$: $v^\dagger \tilde{\PH} v= v^\dagger B^{(\PH)T}(00) v^*= v^\dagger B^{(\PH)}(00) v^* = - v^\dagger B^{(\PH)T}(00) v^*=0$, where the third equation is because $B^{(\PH)}(00)$ is anti-symmetric (\cref{eq:PH-sewing-cond2}).
% In addition, $v$ and $\tilde{\PH}v$ have the same eigenvalue of $\cH(0)$: suppose $\cH(0)v=\lambda v$, then we have $\cH(k_x) \tilde{\PH} v= \tilde{\PH}\cH v= \tilde{\PH} \lambda  v= \lambda \tilde{\PH} v$.

The $\mbb{Z}_2$ invariant $\delta$ is defined such that $\delta=1$ if the Wilson loop bands form a zigzag flow between $k_1=0$ and $k_1=\pi$ - equivalent to a Quantum Spin Hall flow of Kramers paired Wannier centers, and $\delta=0$ otherwise.
Examples of $\delta=1$ and $\delta=0$ \emph{with only $\PH$ symmetry} are  shown in Figs.~\ref{fig:WL}a and~\ref{fig:WL}b, respectively. \cref{fig:WL}a does not contain the $C_{2z}\TRS$ symmetry and is meant to depict the possible cases with only our anti-unitary $\PH$ symmetry. 
Because the degeneracies at $k_1=0,\pi$ are protected by $\PH$, a zigzag flow is stable against adding $\PH$-preserving bands as long as these bands are topologically trivial (they do not exhibit Wilson loop flow themselves) that do not close the gaps between the $2M$ bands and the higher/lower bands
(\cref{fig:WL}a-c). 

In Figs.~\ref{fig:TBG}e and~\ref{fig:TBG}g, we plot the Wilson loop bands of the middle two bands ($\ee_{-1}(\kk),\ee_{1}(\kk)$) of TBG with $\theta=1.05^\circ$ and the Wilson loop bands of the middle ten bands  ($\ee_{-5}(\kk)\cdots \ee_{5}(\kk)$) of TBG with $\theta=0.7^\circ$, respectively. 
Both have the zigzag flow and hence have $\delta=1$.
We do not plot the Wilson loop bands of the middle ten bands of TBG with $\theta=1.05^\circ$ because they have touching points with higher/lower bands at generic momenta (away from high symmetry lines). 

\subsection{Comparison of the \texorpdfstring{$\PH$}{P}-protected topology and \texorpdfstring{$C_{2z}\TRS$}{C2zT}-protected topology}
In Ref. \cite{song_all_2019}, some of the authors of the present work proved that the $C_{2z}\TRS$ symmetry protects the Wilson loop flow for two bands, as shown in \cref{fig:WL}d, where the  crossings at $\lambda=0,\pi$ are protected by $C_{2z}\TRS$.
The Wilson loop flow is characterized by an integer-valued invariant $e_2$: the winding number of a smooth branch of the Wilson loop bands. 
There is a gauge ambiguity for the sign of $e_2$.
For example, the Wilson loop bands in \cref{fig:WL}d has $e_2=1$ if we choose the branch going up to define the winding number and $e_2=-1$ if we choose the branch going down.
$e_2$ is also referred to as the Euler's class \cite{ahn_failure_2019}, as will be briefly introduced in \cref{sec:Chern}. 
With only $C_{2z}\TRS$ symmetry, the flow can be broken by adding two trivial (flat) Wilson loop bands, as shown in \cref{fig:WL}d-f, since the crossings at generic positions - different from $\lambda=0,\pi$- in the Wilson loop spectrum are not protected by $C_{2z}\TRS$. 
After the Wilson loop bands are gapped, one can still define a $C_{2z}\TRS$-protected $\mbb{Z}_2$ invariant through the nested Wilson loop \cite{ahn_failure_2019,song_all_2019}.
Nevertheless, this $C_{2z}\TRS$-protected $\mbb{Z}_2$ invariant does not correspond to Wannier obstruction \cite{ahn_failure_2019,po_faithful_2019}.
Therefore, the topology protected only by $C_{2z}\TRS$ is fragile. Ref.~\cite{song_all_2019} showed that, by adding the unitary particle-hole symmetry $P$, one cannot render the Stiefel–Whitney class \cite{ahn_failure_2019} trivial by adding more bands; however, nontrivial Stiefel-Whitney index does not imply non-Wannierizable bands, and hence \cite{song_all_2019} called the index ``stable'', between quotation marks; this paper removes the quotation marks by proving non-wannieralizability.  

On the contrary, with the $\PH$ symmetry, we cannot break the zigzag flow by adding trivial (non-winding) Wilson loop bands, just like in the Quantum Spin Hall problem.
First, due to the Kramers degeneracy guaranteed by $\PH$, a trivial state must have at least two Wilson loop bands - corresponding to the fact that, with particle-hole symmetry, we must add to the nontrivial bands, generically, two bands - of some energy $\pm E$. 
The two Wilson loop bands are separated at generic $k_1$ but degenerate at $k_1=0,\pi$, as shown in \cref{fig:WL}b.
If we couple such a two-band trivial state to the topological state, the total Wilson loop bands are still gapless (\cref{fig:WL}c) since the degeneracies at $k_1=0,\pi$ are protected.
Therefore, the topology protected by $\PH$ is stable.

If a two-band system has both $C_{2z}\TRS$ and $\PH$ symmetries, the $\mbb{Z}_2$ invariant protected by $\PH$ is given by the parity of $e_2$, \ie $\delta=e_2$ mod 2.
For example, the Wilson loop bands in \cref{fig:TBG}e has $e_2=1$ and $\delta=1$. There is stable topology from $\PH$, in systems with an even number of bands. TBG has even number of bands (as it has to, since $\PH=C_{2z}\TRS P$ implies even number of bands: nonzero energy $E\ne 0$ states come in pairs $\pm E$, while zero energy states have Kramers degeneracy since $\PH^2=-1$); furthermore, these bands exhibit $4l+2$ ($l\in\mbb{N}$) Dirac nodes at zero energy (proved in \cref{sec:no-go}), which will show that TBG is in the topologically nontrivial class of this symmetry.

\subsection{An alternative expression of the \texorpdfstring{$\mathbb{Z}_2$}{Z2} invariant}

We have mentioned that the zigzag flow of the Wilson loop bands protected by $\PH$ is same as the zigzag flow of Wilson loop bands protected by the time-reversal symmetry in 2D Quantum Spin Hall topological insulator \cite{yu_equivalent_2011}. 
Now we show that they are indeed equivalent. 
Suppose $H(\kk)$ have the $\PH$ ($\PH^2=-1$) symmetry, \ie $H(\kk) = -\PH H(-\kk)\PH^{-1}$, then we define the squared Hamiltonian as $H^2(\kk)=H(\kk)\cdot H(\kk)$ such that it commutes with $\PH$, \ie $H^2(\kk) = \PH H^2(-\kk) \PH^{-1}$.
We can regard $\PH$ as a ``time-reversal symmetry'' of $H^2(\kk)$. 
An eigenstate of $H(\kk)$ with the energy $\ee_n(\kk)$ is still an eigenstate of $H^2(\kk)$ but has the squared energy $\ee^2_n(\kk)$. 
States of the $2M$ particle-hole-symmetric bands used to define the Wilson loop (\cref{eq:WL}), \ie $\ket{u_{-M}(\kk)}\cdots \ket{u_{M}(\kk)}$, form the lowest $2M$ bands of the squared Hamiltonian $H^2(\kk)$. 
Thus the Wilson loop operator of the $2M$ particle-hole symmetric bands of $H(\kk)$ is same as the Wilson loop operator of the $2M$ lowest bands of $H^2(\kk)$.
The zigzag flow of the Wilson loop can be equivalently thought as protected by the ``time-reversal symmetry'' of $H^2(\kk)$.
%Therefore, the following two statements are equivalent: (i) $\ket{u_{-M}(\kk)} \cdots \ket{u_{M}(\kk)}$, as eigenstates of $H(\kk)$, have a nontrivial topology protected by $\PH$, (ii) $\ket{u_{-M}(\kk)} \cdots \ket{u_{M}(\kk)}$, as eigenstates of $H^2(\kk)$, form a 2D topological insulator protected by the effective ``time-reversal symmetry'' $\PH$.

\begin{figure}[t]
\centering
\includegraphics[width=0.6\linewidth]{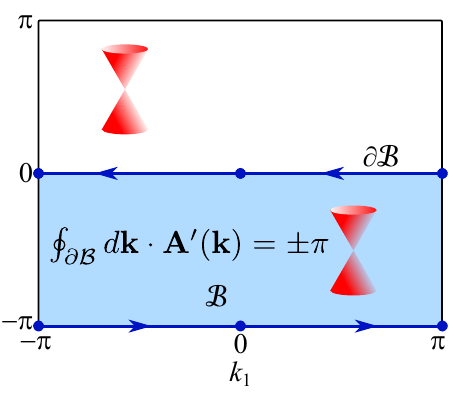}    
\caption{The path used to define the $\mbb{Z}_2$ invariant protected by $\PH$. 
$\mcl{B}=[-\pi,\pi]\otimes[-\pi,0]$ is half of the Brillouin zone. Its boundary $\pt\mcl{B}$ is invariant under the particle-hole symmetry $\PH$ ($\kk\to-\kk$).
With the $C_{2z}\TRS$ symmetry, one Dirac point in $\mcl{B}$ contributes to a $\pi$ Berry's phase along $\partial \mcl{B}$.
}
\label{fig:path}
\end{figure}

The time-reversal-protected $\mathbb{Z}_2$ invariant can be alternatively expressed as a topological obstruction \cite{Fu2006_Z2,fukui2007quantum}.
Consider $2M$ bands $\{\ket{u_n^I(\kk)},\ket{u_n^{II}(\kk)}\, |\, n=1\cdots M\}$ in a time-reversal ($\mcl{T}$) symmetric system that satisfy the gauge condition $\ket{u^{II}_n(-\kk)} = \mathcal{T} \ket{u^{I}_n(\kk)}$, $\ket{u^{I}_n(-\kk)} = -\mathcal{T} \ket{u^{II}_n(\kk)}$,  then the corresponding $\mbb{Z}_2$ invariant is given by
\begin{equation}
\delta = \frac{1}{2\pi} \pare{
    \ointctrclockwise_{\partial \mathcal{B}} d\kk \cdot \mathbf{A}(\kk) - \int_{\mathcal{B}} d^2\kk\, \Omega(\kk)
} \mod 2, \label{eq:Z2-obstruction}
\end{equation}
where $\mcl{B}$ is half of the BZ whose boundary $\pt\mcl{B}$ is $\PH$-invariant, 
\begin{equation}
\mathbf{A}(\kk)=i\sum_{n=1}^M \sum_{a=I,II}\inn{u_n^a(\kk)|\partial_\kk u_n^a(\kk)}
\end{equation}
is the Berry's connection of the considered bands, and $\Omega(\kk)=\pt_\kk \times \mathbf{A}(\kk)$ is the Berry's curvature.
An example of $\mcl{B}$ is shown in \cref{fig:path}.
We regard $\ket{u_{-M}(\kk)}\cdots \ket{u_M(\kk)}$ as the lowest $2M$ bands of $H^2(\kk)$ and $\PH$ the ``time-reversal symmetry'' of $H^2(\kk)$. 
If we impose the gauge $\ket{u_{-n}(-\kk)}=\PH \ket{u_n(\kk)}$ ($n=1\cdots M$), \ie choose the sewing matrix defined in \cref{eq:PH-sewing} as $B^{(\PH)}_{n',n}(\kk) = \delta_{n',-n}\sgn(n)$, then we can regard $\ket{u_n(\kk)}$ and $\ket{u_{-n}(\kk)}$ as $\ket{u^{I}_n(\kk)}$ and $\ket{u^{II}_n(\kk)}$, respectively.
Thus the $\mbb{Z}_2$ invariant of the $2M$ bands of $H(\kk)$ protected by $\PH$ is given by \cref{eq:Z2-obstruction}. 
This expression will be used for one of the ways to prove the symmetry anomaly of $4l+2$ Dirac points in systems with $C_{2z}\TRS$ and $\PH$ symmetries (See \cref{sec:no-go}).

\section{A no-go theorem of two Dirac fermions on lattices with \texorpdfstring{$C_{2z}\TRS$}{C2zT} and \texorpdfstring{$\PH$}{P} symmetries} \label{sec:no-go}

In this section, we will prove that if there are $4l+2$ ($l\in \mbb{N}$) Dirac fermions at zero energy (chemical potential) in a system with $C_{2z}\TRS$ and $\PH$ symmetries, then the $\mbb{Z}_2$ invariant  of the  $2M$ bands (arbitrary $M$) above and below the chemical potential, \ie $\ee_{-M}(\kk)\cdots \ee_{M}(\kk)$, is guaranteed to be 1, provided that the $2M$ bands are gapped from other bands. 
As a consequence, for arbitrary $M$, the $2M$ particle-hole symmetric bands are not Wannierizable. 
That means the $4l+2$ Dirac fermions do not have a lattice support.

Before going into a mathematical proof, we first give an intuitive proof that the Wilson loop of a $C_{2z}\TRS$ and $\PH$  system with $4l+2$ Dirac fermions at zero energy needs to wind. We first assume that we have $2$ bands separate from other bands close to charge neutrality. In \cite{ahn_failure_2019,xie_superfluid_2020} it was shown that the number of Dirac points in half BZ  $\mod 2$ equals the winding of the Wilson  loop. 
For $4l+2$ Dirac nodes in between these two bands, the winding would be odd, as in \cref{fig:TBG}e. Adding non-zero trivial or nontrivial  energy bands to this system would happen in pairs; introducing a set (trivial, due to $C_{2z}T$, which renders Chern numbers to be zero and hence makes any single band topologically trivial) bands at non-zero energy would have its $\PH$ conjugate and appear in numbers $2n$. Introducing nontrivial bands at nonzero energy would mean introducing $2\times2n$ bands into the system, as any possible nontrivial set of bands at a given energy comes as a multiple of $2$.  These bands can introduce only a multiple of $4$ number of Dirac fermions into the system: each set of two separate bands has to have a multiple of $2$ Dirac fermions. From our Quantum Spin Hall (QSH) experience, whatever number of bands we introduce on top of our nontrivial bands with $4l+2$ Dirac fermions cannot change the Wilson loop winding, as we are either adding trivial bands or pairs of nontrivial bands to a QSH system. Hence the winding (of the $4n+2$ Dirac fermion band) is stable to the addition of any bands respecting $C_{2z}\TRS$ and $\PH$. The \emph{only} way the  winding can be interrupted is by the addition of \emph{one} set of $2$ bands with Wilson loop winding to the already existent Wilson loop winding $2$-bands. However, since with $C_{2z} T$ the number of Dirac nodes $\mod 4$ is equal to twice times the winding, this additional one set of $2$-bands would bring about another $4l'+2$ Dirac points so the full system would have a number of Dirac fermions divisible by $4$. Hence a system with $4l+2$ Dirac fermions and $C_{2z}\TRS$ and $\PH$ has to exhibit Wilson loop winding.  

Now, by making use of \cref{eq:Z2-obstruction}, we give another proof that $2M$ bands (gapped from other bands) with $C_{2z}\TRS$ and $\PH$ symmetries that have $4l+2$ Dirac points between $\ee_{-1}(\kk)$ and $\ee_{1}(\kk)$ must have a nontrivial topology.
Due to the $C_{2z}\TRS$ symmetry, the Bloch states satisfy 
\begin{equation}
C_{2z}\TRS \ket{u_n(\kk)} = \sum_{n'}\ket{u_{n'}(\kk)} B^{(C_{2z}\TRS)}_{n'n}(\kk), \label{eq:C2T-sewing}
\end{equation}
where $B^{(C_{2z}\TRS)}_{n'n}(\kk)$ is unitary and called the $C_{2z}\TRS$ sewing matrix.
The summation over $n'$ is limited to values satisfying $\ee_{n'}(\kk)=\ee_{n}(\kk)$.
Substituting this constraint into the definition of the Berry's curvature $\Omega(\kk)$, we find that $\Omega(\kk)=0$ \cite{ahn_failure_2019,xie_superfluid_2020,Slager2019WL}.
Thus we only need to evaluate the first term on the right hand side of \cref{eq:Z2-obstruction}. 
We define 
$\mbf{A}'(\kk)=i \sum_{n=1}^M \inn{u_n(\kk) | \pt_\kk u_n(\kk)}$ for the positive bands and
$\mbf{A}''(\kk)=i \sum_{n=1}^M \inn{u_{-n}(\kk) | \pt_\kk u_{-n}(\kk)}$ for the negative bands.
The total Berry's connection is $\mbf{A}=\mbf{A}'+\mbf{A}''$. 
By imposing the gauge condition $\ket{u_{-n}(-\kk)}=\PH \ket{u_n(\kk)}$ ($n=1\cdots M$) required by \cref{eq:Z2-obstruction}, we find 
{\small
\begin{align}
&\mbf{A}''(\kk) = i \sum_{n=1}^M \inn{u_{-n}(\kk) | \pt_\kk u_{-n}(\kk)} \nono\\
=& i \sum_{n=1}^M \inn{\pt_\kk \PH u_{-n}(\kk) | \PH u_{-n}(\kk)} 
= i \sum_{n=1}^M \inn{\pt_\kk u_{n}(-\kk) |u_{n}(-\kk)} \nono\\
=&- i \sum_{n=1}^M \inn{ u_{n}(-\kk) |\pt_\kk u_{n}(-\kk)} = \mbf{A}'(-\kk),
\end{align}
}%
where we have applied the property of anti-unitary symmetry introduced below \cref{eq:ukukp-P}.
Since the boundary $\pt \mcl{B}$  (\cref{fig:path}) is invariant under $\kk\to -\kk$, the integrals of $\mbf{A}'(\kk)$ and $\mbf{A}''(\kk)$ are equal, \ie
\begin{equation}
\ointctrclockwise_{\pt\mcl{B}} d\kk \cdot \mbf{A}''(\kk) = \ointctrclockwise_{\pt\mcl{B}} d\kk \cdot \mbf{A}'(\kk).\label{eq:A'=A''}
\end{equation}
The $C_{2z}\TRS$ symmetry stabilizes 2D Dirac points \cite{bernevig_topological_2013}, and each Dirac point between the positive bands and the negative bands contribute to a $\pi$ or $-\pi$ Berry's phase of $\mbf{A}'(\kk)$  (\cref{fig:path}). 
Due to the $\PH$ symmetry $\ee_{-n}(-\kk)=-\ee_n(\kk)$, the Dirac points must be equally distributed in $\mcl{B}$ and its complementary set BZ - $\mcl{B}$.
Hence if there are $4l+2$ Dirac points in the BZ, there will be $2l+1$ Dirac points in $\mcl{B}$ and we have $\oint_{\pt\mcl{B}} d\kk \cdot \mbf{A}'(\kk) = (2l+1)\pi$ mod $2\pi$. 
According to \cref{eq:A'=A''}, we have 
\begin{equation}
\ointctrclockwise_{\pt\mcl{B}} d\kk \cdot (\mbf{A}'(\kk) + \mbf{A}''(\kk)) = (4l+2)\pi \mod 4\pi.
\end{equation}
Substituting this equation into \cref{eq:Z2-obstruction} and using the fact that $\Omega(\kk)=0$, we obtain $\delta=1$. 
Thus the presence of $4l+2$ Dirac points in a system with $C_{2z}\TRS$ and $\PH$ symmetries implies a nontrivial topology. 
In contrast to lattice models whose whole bands are trivial, this nontrivial topology is guaranteed by the Dirac points between $\ee_{1}(\kk)$ and $\ee_{-1}(\kk)$ and hence cannot be trivialized by adding higher and lower energy bands (preserving $\PH$). 
Therefore, no matter how many high energy bands are included, as long as they respect $C_{2z} \TRS$ and $\PH$, the considered bands must have have nontrivial topology. 
As will be shown in next paragraph, in a lattice model with a finite number of orbitals per unit cell, the Wilson loop bands of the whole bands must be trivial.
Therefore, $4l+2$ Dirac points cannot be realized in lattice models because the corresponding band structure, no matter how many high and low energy bands are considered, must be topologically nontrivial.

\begin{figure*}
\centering
\includegraphics[width=0.95\linewidth]{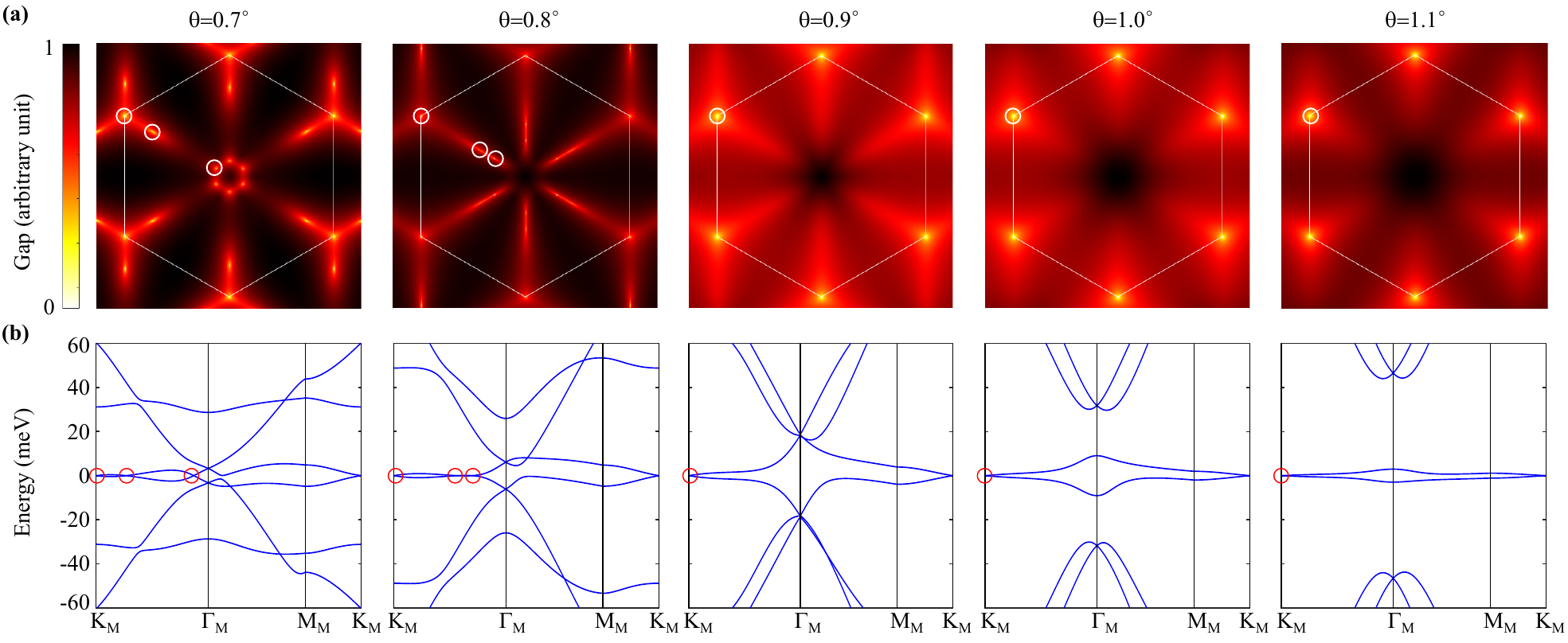}
\caption{The distribution of Dirac points in the Moire Brillouin zone at different twisting angles. (a) The gap between the middle two bands in the Moire Brillouin zone at $\theta=0.7^\circ$, $0.8^\circ$, $0.9^\circ$, $1.0^\circ$, $1.1^\circ$, where the Dirac points along $K_M\Gamma_M$ are marked by the white circles. (b) The corresponding band structures at these twisting angles, where the Dirac points are marked by the red circles. For $\theta\ge 0.9^\circ$, there are two Dirac points locating at $K_M$ and $K_M'$, respectively, in the \MR Brillouin zone. For $\theta\le 0.8^\circ$, there are 14 Dirac points in the \MR Brillouin zone.}
\label{fig:DiracPoint}
\end{figure*}

Here we show that the whole bands of a lattice model must be trivial.
Let the lattice model has $N$ orbitals, then the $U(\kk)$ matrix entering the Wilson loop operator (\cref{eq:WL}) of the whole bands is $U(\kk)=(\ket{u_1(\kk)}\cdots \ket{u_N(\kk)})$. 
By the completeness of all the Bloch states we have $U(\kk)U^\dagger(\kk)=1$.
Thus the Wilson loop operator in \cref{eq:WL} is $W(k_1)=U^\dagger(k_1,0)U(k_1,2\pi) = U^\dagger(k_1,0)V^{(0,2\pi)}U(k_1,0)$, where $V^{(0,2\pi)}$ is the embedding matrix defined in \cref{app:Hk} (with $\GG=2\pi \bb_2$).
Since $U(k_1,0)$ is an $N\times N$ unitary matrix, the eigenvalues of $W(k_1)$ are same as eigenvalues of $V^{(0,2\pi)}$ and hence do not change with $k_1$ and do not wind.

It is worth noting that, in TBG, the symmetry anomaly does not depend on the parameters of the Hamiltonian \cref{eq:Ham}.
In the weak coupling limit ($w_{0}\ll v_F k_D$, $w_1\ll v_Fk_D$), we have two Dirac points at $K_M$ and $K_M'$ in the \MR BZ.
If the $2M$ bands $\ee_{-M}(\kk)\cdots \ee_{M}(\kk)$ are gapped from the other bands, the $2M$ bands must be topological due to correspondence between the number of Dirac points and the $\mbb{Z}_2$ invariant $\delta$.
Tuning the parameters of TBG may couple the $2M$ bands to higher bands $\ee_{M+1}(\kk)\cdots \ee_{M'}(\kk)$ ($M'>M$) and lower bands $\ee_{-M'}(\kk)\cdots \ee_{-M-1}(\kk)$, which are assumed be gapped from $\ee_{M'+1}(\kk)$ and $\ee_{-M'-1}(\kk)$ as we tune the parameters.
In the weak coupling limit, the additional $2M'-2M$ bands must have $\delta=0$ since they do not have Dirac points between $\ee_{-M-1}(\kk)$ and $\ee_{M+1}(\kk)$.
Therefore, after we couple the $2M$ bands to the $2M'-2M$ bands, the $2M'$ bands as a whole will have $\delta=1+0=1$. 
As we tune the parameters, additional Dirac points between $\ee_{1}(\kk)$ and $\ee_{-1}(\kk)$ may be created due to gap closing and reopening between $\ee_{1}(\kk)$ and $\ee_{-1}(\kk)$.
However, the total number of Dirac points between $\ee_{1}(\kk)$ and $\ee_{-1}(\kk)$ must equal to 2 $\mod$ 4, \ie $4l+2$ ($l\in\mbb{N}$), because the topological invariant of the $2M'$ bands is guaranteed to be $\delta=1$.

In \cref{fig:DiracPoint}, we show the evolution of Dirac points with the twisting angle changing from $0.7^\circ$ to $1.1^\circ$. 
For $\theta\in [0.9^\circ,1.1^\circ]$, there are only two Dirac points in the \MR Brillouin zone and they locate at $K_M$ and $K_M'$, respectively. 
When $\theta$ decreases to $0.8^\circ$, two additional Dirac points are generated along the high symmetry line $K_M\Gamma_M$.
Due to the $C_{3z}$ and $\PH$ symmetries, there are twelve Dirac points generated along the equivalent paths of $K_M\Gamma_M$.
Thus for $\theta \in [0.7^\circ, 0.8^\circ]$, there are in total fourteen Dirac points in the Brillouin zone.
Therefore, we always have $4l+2$ ($l\in \mathbb{N}$) Dirac points: for $\theta\in [0.9^\circ,1.1^\circ]$ $l=0$ and for  $\theta\in [0.7^\circ,0.8^\circ]$ $l=3$.

\section{The Chern band basis} \label{sec:Chern}

In this section we show that, if the two bands $\ee_{-1}(\kk)$ and $\ee_{1}(\kk)$ are gapped from other bands, we can recombine them as two Chern bands with Chern numbers $e_2$ and $-e_2$, with $e_2$ being  the Euler's class \cite{ahn_failure_2019,ahn_band_2018,Slager2020Euler,wu2019non} (or, equivalently, the Wilson loop  winding number protected by $C_{2z}\TRS$ \cite{xie_superfluid_2020}).
(In TBG, the Chern numbers given by $\pm e_2$ are also equal to the $e_Y=\pm1$ index defined in Ref. \cite{ourpaper3}, which, in a certain gauge, represents the eigenvalue of the Pauli $y$ matrix in the 2-dimensional space of $n=\pm1$ band indices.)
%\textbf{BAB: mention that $e_2$ is different from $e_\pm$ used in our interacting paper }

In order to introduce the Chern band basis, we first introduce the definition of Euler's class $e_2$.
(We refer the readers to Refs. [\onlinecite{ahn_failure_2019}, \onlinecite{ahn_band_2018}, \onlinecite{xie_superfluid_2020}] for more details.)
Suppose the two bands $\ee_1(\kk)$ and $\ee_{-1}(\kk)$ are gapped from other bands. 
Then, at $\kk$ away from Dirac points between the two bands, the $C_{2z}\TRS$ operator leaves each band unchanged up to a phase factor.
In other words, the $C_{2z}\TRS$ sewing matrix (\cref{eq:C2T-sewing}) is diagonal at these $\kk$.
Hence in general, the $C_{2z}\TRS$ symmetry acts on the Bloch states as $C_{2z}\TRS \ket{u_n(\kk)} = \ket{u_n(\kk)} e^{i\theta_n(\kk)}$ ($n=\pm1$), with $\theta_n(\kk)$ being the phase factors.
According to Ref. \cite{ahn_failure_2019}, it follows that the non-Abelian Berry's connection of the two bands at $\kk$ away from Dirac points takes the form 
\begin{align}
& \cA_{nn'}(\kk) = i\inn{u_{n}(\kk)|\pt_\kk u_{n'}(\kk)} \nono\\
=& \begin{pmatrix}
-\frac12 \partial_\kk \theta_1(\kk) & -i\aa(\kk) e^{i\frac{\theta_1(\kk)-\theta_{-1}(\kk)}2}\\
i\aa(\kk) e^{i\frac{\theta_{-1}(\kk)-\theta_1(\kk)}2} & -\frac12 \partial_\kk \theta_{-1}(\kk)
\end{pmatrix}_{nn'}.\label{eq:A-C2T}
\end{align}
$\aa(\kk)$ is a gauge invariant quantity up to a global ambiguity of $\pm$ sign.
The Euler's class is given by 
{\small 
\begin{equation}
e_2 = \frac1{2\pi} \sum_{i} \ointctrclockwise_{\partial D_i} d\kk \cdot \aa(\kk)
    = \frac1{2\pi} \int_{{\rm BZ'}} d^2\kk\ f(\kk) \quad 
\in \mathbb{Z}.
\end{equation}
}%
Here $i$ indexes the Dirac points in the BZ, $D_i$ is a sufficiently small region covering the $i$-th Dirac point, ${\rm BZ'}={\rm BZ} - \sum_i D_i$, and $f(\kk)=\pt_\kk\times\aa(\kk)$.

In the above we have assumed that the $\ket{u_n(\kk)}$ is smooth over the Brillouin zone {\it except} at the Dirac points. 
\cref{eq:A-C2T} is valid only in this gauge.
In this gauge $\theta_n(\kk)$ is necessarily k-dependent if there exist Dirac points between the $n$th band and other bands.
Since each Dirac point contributes to a $\pi$ Berry's phase, there must be $2\ointctrclockwise_{\partial D_i} d\kk \cdot  \boldsymbol{\mathcal{A}}_{nn}(\kk) = \ointctrclockwise_{\partial D_i} d\kk \cdot \partial_\kk \theta_n(\kk) = 2\pi$ mod $4\pi$.
Thus $\theta_n(\kk)$ must wind odd times around a Dirac point.

We introduce the two Chern band basis as 
\begin{equation}
\ket{v_\pm(\kk)} = \frac{1}{\sqrt2}( e^{i\frac{\theta_1(\kk)}2} \ket{u_1(\kk)} \pm i e^{i\frac{\theta_{-1}(\kk)}2} \ket{u_{-1}(\kk)}). \label{eq:Chern-basis}
\end{equation}
There are two ambiguities in the above equation: (i) There is an ambiguity of the two branches of $\frac{\theta_n}2$, \ie $\frac{\theta_n}2$ and $\frac{\theta_n}2+\pi$.
(ii) At the Dirac points, where the two bands are degenerate, there is an ambiguity of choosing $u_{1}(\kk)$ and $u_{-1}(\kk)$.
Replacing $\theta_1(\kk)/2$ by $\theta_1(\kk)/2+\pi$ or replacing $\theta_{-1}(\kk)/2$ by $\theta_{-1}(\kk)/2+\pi$ will interchange $\ket{v_+(\kk)}$ with $\ket{v_-(\kk)}$. 
Similarly, interchanging $\ket{u_{1}(\kk)}$ and $\ket{u_{-1}(\kk)}$ at the Dirac points will also interchange  $\ket{v_+(\kk)}$ with $\ket{v_-(\kk)}$ at the Dirac points. 
To solve these ambiguities, as detailed in \cref{app:Chern-gauge}, we require that the Berry's curvatures of $\ket{v_+(\kk)}$ and $\ket{v_-(\kk)}$ to be continuous, or, equivalently, 
\begin{equation}
\lim_{\qq\to 0} |\inn{v_{m'}(\kk+\qq) | v_{m}(\kk) }| = \delta_{m'm}, \label{eq:Chern-band-gauge}
\end{equation}
where $m,m'=\pm$. 
Using \cref{eq:A-C2T,eq:Chern-basis}, we can calculate the non-Abelian Berry's connection on the basis $\ket{v_\pm(\kk)}$ at $\kk$ away from the Dirac points.
We obtain
\begin{equation}
\cA_{mm'}'(\kk) = i\inn{v_{m}(\kk)|\pt_\kk v_{m'}(\kk)} = 
\begin{pmatrix}
\aa(\kk) & 0 \\ 0 & -\aa(\kk)
\end{pmatrix}_{mm'},
\end{equation}
and hence 
\begin{equation}
\cF'_{mm'}=-[\pt_{k_x}-\cA_{x}',\pt_{k_y}-\cA_{y}']_{mm'}=
\begin{pmatrix}
f(\kk) & 0 \\ 0 & -f(\kk)
\end{pmatrix}_{mm'},
\end{equation}
for $\kk$ not at the Dirac points. 
Therefore, if the Berry's curvature does not diverge at Dirac points, which is true as shown in next paragraph, the Chern numbers of the states $\ket{v_\pm(\kk)}$ are 
\begin{equation}
C_\pm = \pm\frac1{2\pi} \int d^2\kk\ f(\kk) = \pm e_2. 
\end{equation}

To conclude this section, we show that in the chiral limit $w_0=0$ the Chern band basis can be chosen as the eigenstates of the chiral symmetry $C$ (\cref{eq:chiral}).
We define the sewing matrix of $C$ as
\begin{equation}
C \ket{u_n(\kk)} = \sum_{n'} \ket{u_{n'}(\kk)} B^{(C)}_{n'n}(\kk), \label{eq:sewing-S}
\end{equation}
where the summation over $n'$ satisfies $\ee_{n'}(\kk)=-\ee_n(\kk)$. 
For the TBG Hamiltonian \cref{eq:Ham}, the $C_{2z}\TRS$ and $C$ operators are $\sigma_x K$ and $\sigma_z$, respectively. 
Thus we have the algebra $C^2=1$ and $\{C,C_{2z}\TRS\}=0$ and hence 
\begin{equation}
[B^{(C)}(\kk)]^2=1, \label{eq:BSeq1}
\end{equation}
\begin{equation}
B^{(C_{2z}\TRS)}(\kk) B^{(C)*}(\kk)+B^{(C)}(\kk) B^{(C_{2z}\TRS)}(\kk)=0. \label{eq:BSeq2}
\end{equation}
As discussed at the beginning of this section, at $\kk$ not at the Dirac points, we have $B^{(C_{2z}\TRS)}_{n'n}(\kk)=\delta_{n'n} e^{i\theta_n(\kk)}$. 
Then the solution of $B^{(C)}(\kk)$ is 
\begin{equation}
B^{(C)}(\kk) = \pm \begin{pmatrix}
0 & -i e^{i\frac{\theta_1(\kk)-\theta_{-1}(\kk)}2}\\
i e^{i\frac{\theta_{-1}(\kk)-\theta_{1}(\kk)}2} & 0
\end{pmatrix}.
\end{equation}
The $\pm$ sign cannot be determined by solving \cref{eq:BSeq1,eq:BSeq2}.
In practice, one should evaluate \cref{eq:sewing-S} to determine the $\pm$ sign for given $\ket{u_{\pm1}(\kk)}$.
We find that the Chern band basis \cref{eq:Chern-basis} diagonalizes $B^{(C)}(\kk)$. 
Below \cref{eq:Chern-basis} we have discussed the ambiguity of choosing $\ket{v_\pm(\kk)}$ and we have imposed \cref{eq:Chern-band-gauge} to fix this ambiguity.
This ambiguity of \cref{eq:Chern-basis} can be alternatively solved by choosing $\ket{v_\pm(\kk)}$ as the eigenstates of $C$ with the eigenvalues $\pm1$, respectively. 
This choice automatically satisfies \cref{eq:Chern-band-gauge} since the states with different chiral eigenvalues are orthogonal, \ie $\inn{v_-(\kk)|v_+(\kk')}=\inn{v_-(\kk)|C^\dagger C|v_+(\kk')}=-\inn{v_-(\kk)|v_+(\kk')}=0$ for arbitrary $\kk$ and $\kk'$.

The Chern band basis in Eq. (\ref{eq:Chern-basis}) can also be equivalently defined through the Wilson loop method \cite{kang_nonabelian_2020,hejazi2020hybrid}. Besides, in the chiral limit, the Chern band basis we defined is equivalent to that defined in \cite{bultinck_ground_2020}.

\section{Perfect metal phase of twisted bilayer graphene in the second chiral limit} \label{sec:S2}

In our article Ref.~\cite{ourpaper3}, we consider the opposite limit of the usual chiral limit: Instead of letting $w_0=0$, we take $w_1$ to be zero.
When $w_1=0$, the model \cref{eq:Ham} has another chiral symmetry $C'=\tau_z\sigma_z$ acting on the Hamiltonian as $C' H(\rr) C^{\prime\dagger} = -H(\rr)$.
Thus we call this limit as the second chiral limit. 
As discussed in \cref{sec:TBG-review}, $w_0$ and $w_1$ are the interlayer couplings contributed mainly by the AA and AB/BA regions, respectively; thus the second chiral limit can be (approximately) realized if the layer distance in the AA region is smaller than the layer distance in the AB and BA regions (shorter distance means stronger coupling).
Such a configuration would be different from the corrugation predicted by the first principle calculations \cite{Uchida_corrugation,Wijk_corrugation,dai_corrugation,jain_corrugation}, where the distance in the AA region is larger.
Nevertheless, the second chiral limit might could potentially be engineered by putting the TBG on certain substrate, and it represents an interesting interacting limit \cite{ourpaper3}.
We are mainly interested in the novel electronic band structure of TBG in the second chiral limit and hence we leave the material realization of the second chiral limit for future study.

\begin{figure}
\centering
\includegraphics[width=1\linewidth]{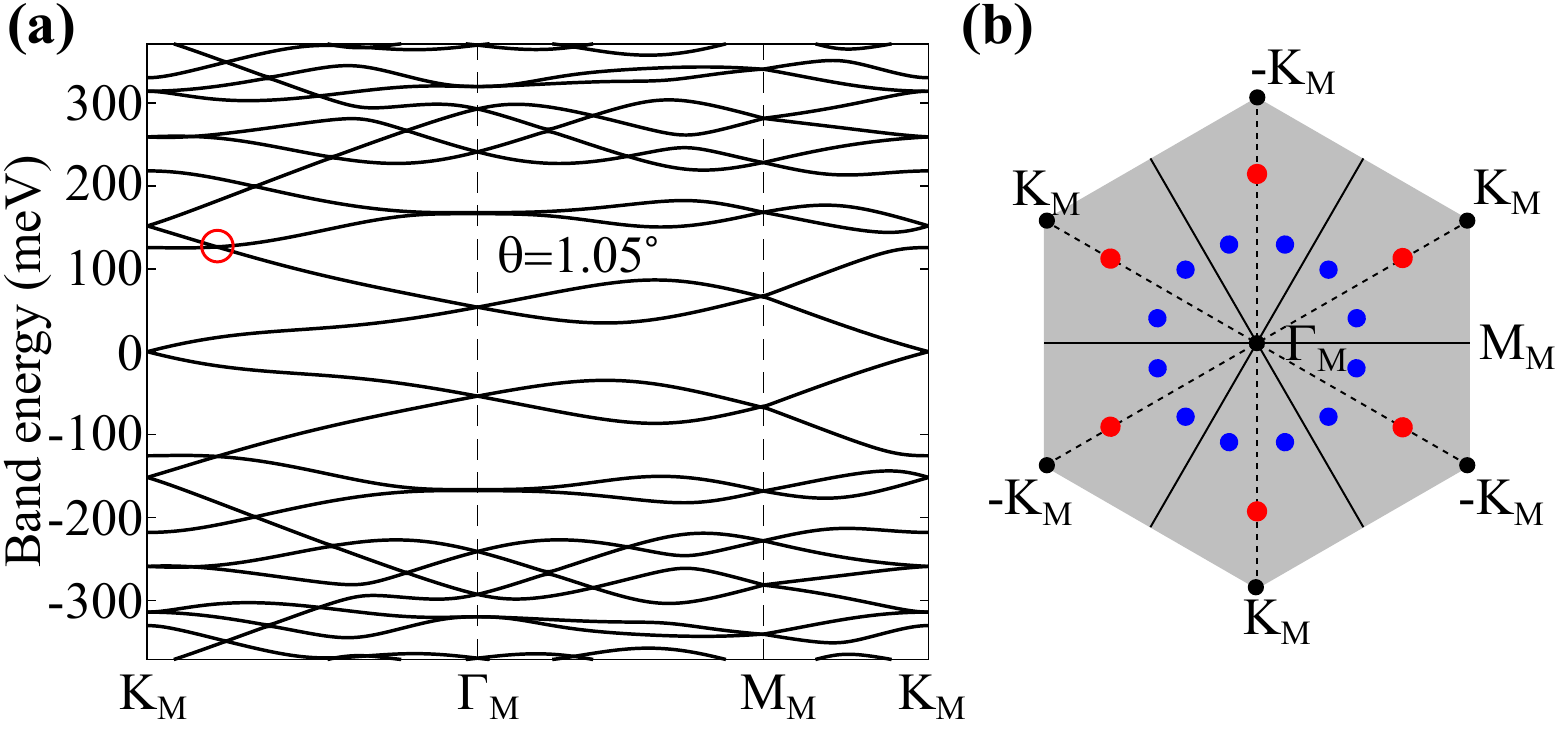}
\caption{Perfect metal phase of twisted bilayer graphene in the second chiral limit ($w_1=0$).
\textbf{(a)} The band structure at $\theta=1.05^\circ$ with the parameters $v_F=5.944{\rm eV \cdot \mathring{ A} }$, $|K|=1.703\mathring{\rm A}^{-1}$, $w_1=0$, $w_0=77{\rm meV}$.
\textbf{(b)} The Brillouin zone of the twisted bilayer graphene. The solid black lines represent the $C_{2x}$-axis and its conjugations under $C_{3z}$, the dashed black lines represent the effective mirror symmetry $M_x=C_{2x}I$ and its conjugations under $C_{3z}$.  
The red dots represent Dirac points in the mirror lines. 
The blue dots represent Dirac points at generic momenta.
}
\label{fig:second}
\end{figure}

We find that, in the second chiral limit, the $n$-th positive (negative) band is always connected to the $(n+1)$-th positive (negative) band.
As the first positive band and the first negative band are connected through the Dirac points, the {\it whole} bands are all connected, as shown in \cref{fig:second}a.
The phase with all bands connected is referred to as the perfect metal \cite{Mora2019-graphene} in trilayer systems, where the number of Dirac nodes is odd. In the current case, we also find this ``perfect metal" in even number of Dirac node systems with the special chiral symmetry of the second chiral limit.  

The perfect metal phase is protected by $C_{2z}\TRS$, $P$, and $C'$. The new chiral symmetry $C'$ has a strange group algebra as it anticommutes with $\TRS$ and with $P$ \cite{ourpaper4}.
We define the product of $P$ and $C'$ as an effective inversion symmetry $I=PC'=\tau_x\sigma_z$.
It commutes with the Hamiltonian, \ie $H(-\rr)=I H(\rr) I^\dagger$ and $H(-\kk)=I H(\kk) I^\dagger$ accordingly.
%In the following, we will make use of $I$ instead of $C'$ for convenience.
The effective inversion operator satisfies the algebra
\begin{equation}
\{C_{2z}\TRS,I\}=0,\qquad \{P,I\}=0,\qquad I^2=1.
\end{equation}
We first show that $C_{2z}\TRS$ and $I$ protect double degeneracies at $I$-invariant momenta.
(In TBG, the $I$-invariant momenta are $\Gamma_M$ and the three equivalent $M_M$.)
Since the Hamiltonian at an $I$-invariant momentum commutes with $I$, the Bloch states at this momentum must form eigenstates of $I$.
Suppose $\ket{u}$ is such an eigenstate with $I$ eigenvalue 1, then we can show that $C_{2z}\TRS \ket{u}$ must have the opposite $I$ eigenvalue -1 due to the anti-commutation between $C_{2z}\TRS$ and $I$. 
Therefore $\ket{u}$ and $C_{2z}\TRS\ket{u}$ form a doublet that has opposite $I$ eigenvalues.
This explains the double degeneracies at $\Gamma_M$ and $M_M$ shown in \cref{fig:second}a. 

Next we prove that, for arbitrary even $M$, the $M$-th positive band is connected to the $(M+1)$-th positive bands through $4l+2$ ($l\in\mbb{N}$) Dirac points.
(For odd $M$, we know by counting - see \cref{fig:second} - that the $M$-th band is connected to the $(M+1)$-th band through the double degeneracies at the $I$-invariant momenta.)
We only need to prove for the situation where the four bands $\ee_{M-1}(\kk)$, $\ee_{M}(\kk)$, $\ee_{M+1}(\kk)$, $\ee_{M+2}(\kk)$ do {\it not} form four-fold degeneracies at high symmetry momenta since otherwise $\ee_{M}(\kk)$ is already connected to $\ee_{M+1}(\kk)$.
(As shown in \cref{fig:second}a, we also do not observe four-fold degeneracies at high symmetry momenta.)
We assume there are in total $n_D$ Dirac points between the first $M$ positive bands $\ee_1(\kk)\cdots \ee_{M}(\kk)$ and the other bands.
The $n_D$ Dirac points can appear above the $M$-th band, \ie between $\ee_M(\kk)$ and $\ee_{M+1}(\kk)$, or below the first band, \ie between $\ee_1(\kk)$ and $\ee_{-1}(\kk)$.
According to the $I$ symmetry, half of the BZ ($\mcl{B}$) must have $n_D/2$ Dirac points.
(The choice of $\mcl{B}$ is not unique. An example is shown in \cref{fig:path}.)
With the $C_{2z}\TRS$ symmetry, the number of Dirac points in $\mcl{B}$ is related to the Berry's phase surrounding $\mcl{B}$ as \cite{bernevig_topological_2013}
\begin{equation}
\frac{n_D}2 = \frac1{\pi} \ointctrclockwise_{\pt\mcl{B}} d\kk \cdot \mbf{A}'(\kk) \mod 2, \label{eq:nD-A}
\end{equation}
where $\mbf{A}'(\kk)=\sum_{n=1}^M i\inn{u_n(\kk)|\pt_\kk u_n(\kk)}$ is the Berry's connection of the first $M$ positive bands.
In presence of the effective inversion symmetry $I$, the right hand side of the above equation is determined by the $I$ eigenvalues as \cite{fang_bulk_2012,Hughes2011Inv,Ashvin2012Inv,Aris2014WL} 
\begin{equation}
\exp\pare{i \ointctrclockwise_{\pt\mcl{B}} d\kk \cdot \mbf{A}'(\kk)} = \prod_{{\cal K}} \prod_{n=1}^M \xi_{{\cal K},n}, \label{eq:A-xi}
\end{equation}
where ${\cal K}$ indexes the four $I$-invariant momenta, and $\xi_{{\cal K},n}$ is the $I$ eigenvalue of the $n$-th positive band at the momentum ${\cal K}$.
As discussed in the last paragraph, each doublet at an $I$-invariant momentum has opposite $I$ eigenvalues. 
Thus, there are equal number of $I$ eigenvalues 1 and $-1$ at the $I$-invariant momenta; since the total number of states at the four $I$-invariant momenta is $4M$, there are $2M$ eigenvalues with $I=+1$ and $2M$ eigenvalues with $I=-1$. 
Hence the right hand side of \cref{eq:A-xi} is 1 and we have $\ointctrclockwise_{\pt\mcl{B}}  d\kk \cdot \mbf{A}'(\kk) =0$ mod $2\pi$. 
According to \cref{eq:nD-A}, the total number of Dirac points is a multiple of 4, \ie $n_D=0$ mod 4. 
As we have proved in \cref{sec:no-go}, there must be $4l'+2$ ($l'\in\mbb{N}$) Dirac points between $\ee_{1}(\kk)$ and $\ee_{-1}(\kk)$, then the number of Dirac points between $\ee_M(\kk)$ and $\ee_{M+1}(\kk)$ is $n_D-4l'-2=4l+2$ with $l=n_D/4 - l' -1 \in \mbb{N}$.
Thus the $M$-th positive band is always connected to the $(M+1)$-th positive band through $4l+2$ Dirac points.
According to the particle-hole symmetry $P$, the $M$-th negative band is also connected to the $(M+1)$-th negative band through $4l+2$ Dirac points.
Therefore, the whole set of bands in the system will be connected.

In general, the $4l+2$ Dirac points between the $M$-th band the $(M+1)$-th band can be located anywhere in the BZ.
However, with the $C_{3z}$ and the $C_{2x}$ symmetries of TBG, at least some of the $4l+2$ Dirac points must locate at high symmetry point or along high symmetry lines of the BZ.
We prove this statement by contradiction.
The unitary point group of TBG is generated by $C_{3z}$, $C_{2x}$, and the effective inversion $I$ and hence is isomorphic to the point group $D_{3d}$, which has 12 elements in total.
If all the Dirac points between the $M$-th band the $(M+1)$-th band are located at generic momenta, then the number of Dirac points would be a multiple of 12, as represented by the blue dots in \cref{fig:second}b, leading to a contradictory with the $4l+2$ Dirac points.
Therefore, there must be 2 (modulo 4) Dirac points at the high symmetry points or along the high symmetry lines.
As a consequence, the entire set of bands of TBG in the second chiral limit must be connected along the high symmetry lines. 
For example, as shown in \cref{fig:second}a, there is a crossing between the 2nd and 3rd bands in the high symmetry line $\Gamma_M-K_M$. 
(This crossing is protected by the effective mirror symmetry $M_x=C_{2x}I$.)
Under the actions of $C_{2x}$ and $C_{3z}$, there are in total six symmetry counterparts of this crossing point (including itself).
Thus the number of Dirac points is consistent with $4l+2$ with $l=1$.

%It worth noting that, with $I$, we can define an effective mirror symmetry $M_x=C_{2x}I$.
%The high symmetry points and lines in the BZ are shown in \cref{fig:second}c.

\section{Conclusions}\label{sec:conclusion}

In this work, we showed that even the simple, well studied BM TBG model still has several surprises related to the deep physics that it describes. We have proved that the band structure in a single graphene valley of TBG is anomalous, \ie does not have lattice support that respects the $C_{2z}\TRS$ and $\PH$ symmetries.
The anomaly manifests as (i) a $\mbb{Z}_2$ nontrivial topology protected $\PH$ of the $2M$ bands $\ee_{-M}(\kk)\cdots\ee_{M}(\kk)$ for arbitrary $M$, provided that the $2M$ bands are gapped from other bands, (ii) $4l+2$ ($l\in\mbb{N}$) Dirac points between $\ee_{-1}(\kk)$ and $\ee_{1}(\kk)$. 
In the second chiral limit ($w_1=0$), the anomaly manifests as (iii) a perfect metal phase where all the bands are connected.

As a consequence of the symmetry anomaly, a faithful description of TBG that respects all the symmetries of TBG, including $\PH$, \emph{is forced to}  adopt a momentum space formalism.
Any tight-binding description \cite{po_faithful_2019,kang_symmetry_2018,koshino_maximally_2018,bultinck_ground_2020,Wilson2020TBG} of TBG with finite number of orbitals must break at least one of the $C_{2z}\TRS$ and $\PH$ symmetries (or the valley symmetry if the tight-binding model mix the two graphene valleys of TBG).
In the other works of our series on TBG \cite{ourpaper1,ourpaper3,ourpaper4,ourpaper5,ourpaper6}, the interacting physics is studied using a momentum space formalism. 

\begin{acknowledgments}
We thank Aditya Cowsik and Fang Xie for valuable discussions.
This work was supported by the DOE Grant No. DE-SC0016239, the Schmidt Fund for Innovative Research, Simons Investigator Grant No. 404513, the Packard Foundation, the Gordon and Betty Moore Foundation through Grant No. GBMF8685 towards the Princeton theory program, and a Guggenheim Fellowship from the John Simon Guggenheim Memorial Foundation. Further support was provided by the NSF-EAGER No. DMR 1643312, NSF-MRSEC No. DMR-1420541 and DMR-2011750, ONR No. N00014-20-1-2303, Gordon and Betty Moore Foundation through Grant GBMF8685 towards the Princeton theory program, BSF Israel US foundation No. 2018226, and the Princeton Global Network Funds. 
\end{acknowledgments}

\bibliography{ref,HexalogyInternalRefs}
\clearpage

\appendix
\section{Hamiltonian of twisted bilayer graphene in momentum space} \label{app:Hk}

\subsection{The Hamiltonian}
\begin{figure}[t]
\centering
\includegraphics[width=\linewidth]{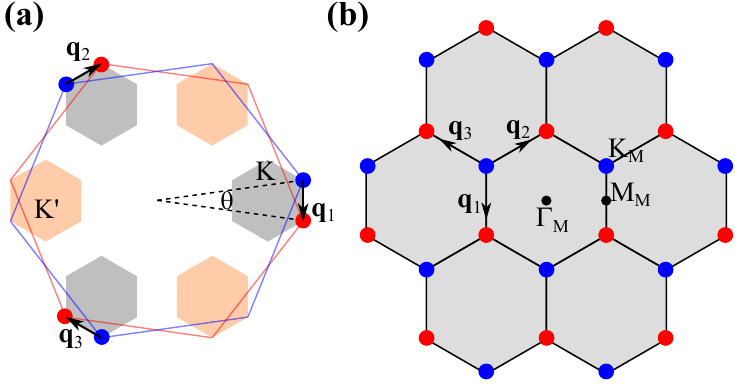}
\caption{The $\QQ$-lattice for the momentum space Hamiltonian of twisted bilayer graphene. 
\textbf{(a)} The blue and red hexagons represent the Brillouin zones of the top layer and the bottom layer, respectively. The blue and red dots represent the positions of Dirac points of the two layers in the graphene valley $K$, respectively. 
\textbf{(b)} The $\QQ$ lattice formed by adding $\qq_{1,2,3}$ iteratively. At each blue dot a plane-wave state from the top layer is assigned, and at each red dot a plane-wave state from the bottom layer is assigned. 
}
\label{fig:QQ}
\end{figure}

Here we briefly introduce the momentum space Hamiltonian $H(\kk)$ corresponding to \cref{eq:Ham}.
Readers may refer to the supplementary materials of Ref. \cite{song_all_2019} for more details.
The basis of $H(\kk)$ is $\ket{\phi_{\QQ,\alpha}(\kk)}$, where $\kk$ is a momentum in the \MR BZ, $\QQ$ is a point in the lattice shown in \cref{fig:QQ}, and $\alpha=1,2$ is the sublattice index of graphene.
There are two types of $\QQ$ lattices: the blue lattice $\mcl{Q}_{T}$ and the red lattice $\mcl{Q}_{B}$.
For $\QQ \in \mcl{Q}_T$, the basis $\ket{\phi_{\QQ,\alpha}(\kk)}$ is a plane-wave state from the top layer
\begin{equation}
\ket{\phi_{\QQ,\alpha}(\kk)} = \frac{1}{\sqrt{N}} \sum_{\RR} e^{i(\RR+\tt_\alpha)\cdot(\kk-\QQ)} \ket{\RR+\tt_\alpha}, \label{eq:Hk-basis1}
\end{equation}
where $N$ is the number of lattices in the top layer graphene, $\RR$ indexes all the lattices of the top layer graphene, $\tt_\alpha$ is the sublattice vector of the top layer graphene, $\ket{\RR+\tt_\alpha}$ is the atomic orbital at $\RR+\tt_\alpha$.
For $\QQ \in \mcl{Q}_B$, the basis $\ket{\phi_{\QQ,\alpha}(\kk)}$ is a plane-wave state from the bottom layer
\begin{equation}
\ket{\phi_{\QQ,\alpha}(\kk)} = \frac{1}{\sqrt{N}} \sum_{\RR'} e^{i(\RR'+\tt_\alpha')\cdot(\kk-\QQ)} \ket{\RR'+\tt_\alpha'},\label{eq:Hk-basis2}
\end{equation}
where $N$ is the number of lattices in the bottom layer graphene (same as the one in top layer), $\RR'$ indexes all the lattices of the bottom layer graphene, $\tt_\alpha'$ is the sublattice vector of the top layer graphene, $\ket{\RR'+\tt_\alpha'}$ is the atomic orbital at $\RR'+\tt_\alpha'$.
The Hamiltonian (\cref{eq:Ham}) on the basis $\ket{\phi_{\QQ,\alpha}(\kk)}$ is given by 
\begin{align}
H_{\QQ,\QQ'}(\kk) =& v_F \delta_{\QQ,\QQ'} (\kk-\QQ) \cdot\bsigma -\frac{\theta}2 \zeta_\QQ \delta_{\QQ,\QQ'} (\kk-\QQ) \times\bsigma  \nono\\
+ &\sum_{j=1}^3 (\delta_{\QQ'-\QQ,\qq_j}+\delta_{\QQ-\QQ',\qq_j}) T_j, \label{eq:Hk}
\end{align}
where $\bsigma=(\sigma_x,\sigma_y)$ and $T_j$ ($j=1,2,3$) are two-by-two matrices in the sublattice space, $\zeta_{\mathbf{Q}}=1$ for $\QQ \in \mcl{Q}_T$ and $\zeta_{\mathbf{Q}}=-1$ for $\QQ\in\mcl{Q}_B$. 

\subsection{The periodicity of Bloch states}

An eigenstate of \cref{eq:Ham} at a given momentum $\kk$ can be written as linear combination of the Bloch basis as $\ket{\psi_{n}(\kk)} = \sum_{\QQ,\alpha} \ket{\phi_{\QQ,\alpha}(\kk)} u_{\QQ\alpha, n}(\kk)$. 
Here $n$ is the band index. 
It should be noticed that the basis states in \cref{eq:Hk-basis1,eq:Hk-basis2} are {\it not} periodic in the \MR BZ since $\ket{\phi_{\QQ,\alpha}(\kk+\GG)} = \ket{\psi_{\QQ-\GG,\alpha}(\kk)}$ by definition. 
In order for the Bloch state $\ket{\psi_{n}(\kk)}$ to be periodic in the \MR BZ, \ie $\ket{\psi_{n}(\kk+\GG)} = \ket{\psi_{n}(\kk)}$, $u_{\QQ,\alpha} (\kk)$ should satisfy $
u_{\QQ,\alpha}(\kk+\GG) = u_{\QQ-\GG,\alpha}(\kk)$. 
We introduce the embedding matrix 
\begin{equation}
V^\GG_{\QQ,\QQ'} = \delta_{\QQ-\GG,\QQ'}
\end{equation}
such that we can write the periodicity of Bloch states as 
\begin{equation}
\ket{u_n(\kk+\GG)} = V^\GG \ket{u_n(\kk)},
\end{equation}
where $\ket{u_n(\kk)} = (u_{\QQ_1,1,n}(\kk), u_{\QQ_1,2,n}(\kk), u_{\QQ_2,1,n}(\kk)\cdots)^T$. While exact Bloch periodicity requires that the cutoff in the lattice $\QQ$ be large, we have showed - around the first magic angle -  \cite{ourpaper1} that we can obtain machine precision accuracy in the first \MR BZ by taking a small cutoff in $\QQ$

\subsection{Symmetry operators in the momentum space}

The crystalline symmetry group of the single valley Hamiltonian is the magnetic space group $P6^{\prime}2^{\prime}2$ (\#177.151 in the BNS setting \cite{Bilbao-MSG}). The generators of this group are:
The $C_{3z}$ symmetry 
\begin{equation}
H\left(C_{3z}\mathbf{k}\right)=D \left(C_{3z}\right)H\left(C_{3z}\mathbf{k}\right)D^\dagger \left(C_{3z}\right)
\end{equation}
where $D_{\mathbf{Q}^{\prime},\mathbf{Q}}\left(C_{3z}\right)=e^{i\frac{2\pi}{3}\sigma_{z}}\delta_{\mathbf{Q}^{\prime},C_{3z}\mathbf{Q}}$.
The $C_{2x}$ symmetry
\begin{equation}
H\left(C_{2x}\mathbf{k}\right)=D\left(C_{2x}\right)H\left(C_{2x}\mathbf{k}\right)D^{\dagger}\left(C_{2x}\right)
\end{equation}
where $D_{\mathbf{Q}^{\prime},\mathbf{Q}}\left(C_{2x}\right)=\sigma_{x}\delta_{\mathbf{Q}^{\prime},C_{2x}\mathbf{Q}}$,
and $C_{2x}\mathbf{q}_{1}=-\mathbf{q}_{1}$.
The $C_{2z}\TRS$ symmetry
\begin{equation}
H\left(\mathbf{k}\right)=D\left(C_{2z}\TRS\right)H_{\mathbf{Q},\mathbf{Q}^{\prime}}^{*}\left(\mathbf{k}\right)D^{T}\left(C_{2z}\TRS\right)
\end{equation}
where $D_{\mathbf{Q}^{\prime},\mathbf{Q}}\left(C_{2z}T\right)=\sigma_{x}\delta_{\mathbf{Q}^{\prime},\mathbf{Q}}$.
It should be noticed that all  rotations of momenta here are with respect to the $\Gamma_M$ point of the \MR BZ.

When the second term in \cref{eq:Hk} is negligible, $H(\kk)$ has an emergent unitary particle-hole symmetry
\begin{equation}
H\left(-\mathbf{k}\right) = -D\left(P\right)H\left(\mathbf{k}\right)D^{\dagger}\left(P\right) 
\end{equation}
where $D_{\mathbf{Q}^{\prime},\mathbf{Q}}\left(P\right)=\delta_{\mathbf{Q}^{\prime},-\mathbf{Q}}\zeta_{\mathbf{Q}}$, and $\zeta_{\mathbf{Q}}=1$ for $\QQ \in \mcl{Q}_T$ and $\zeta_{\mathbf{Q}}=-1$ for $\QQ\in\mcl{Q}_B$. 
The anti-unitary particle-hole symmetry $\PH= P C_{2z}\TRS$ in momentum space is 
\begin{equation}
H\left(-\mathbf{k}\right) = -D(\PH) H^*(-\kk) D^T(\PH),
\end{equation}
where $D_{\QQ,\QQ'}(\PH) = \sigma_x \delta_{\mathbf{Q}^{\prime},-\mathbf{Q}}\zeta_{\mathbf{Q}}$.
When $w_0=0$, $H(\kk)$ has an emergent chiral symmetry
\begin{equation}
H\left(\mathbf{k}\right) = -D^{\dagger}(C) H(\kk) D(C), 
\end{equation}
where $D_{\QQ,\QQ'}(C)=\sigma_z \delta_{\QQ,\QQ'}$. 
When $w_1=0$, $H(\kk)$ has an another emergent chiral symmetry (the second chiral symmetry)
\begin{equation}
H\left(\mathbf{k}\right) = -D^{\dagger}(C') H(\kk) D(C'), 
\end{equation}
where $D_{\QQ,\QQ'}(C')=\sigma_z \delta_{\QQ,\QQ'} \zeta_\QQ$. 

The embedding matrix transform under the unitary operators ($g=C_{3z},C_{2x},P,C,C'$) as
\begin{equation}
D(g) V^\GG D^\dagger(g) = V^{g\GG}. \label{eq:VD1}
\end{equation}
For the unitary operator $C_{2z}\TRS$, we have
\begin{equation}
D(C_{2z}\TRS) V^{\GG*} D^T(C_{2z}\TRS) = V^{\GG}. \label{eq:VD2}
\end{equation}
One can verify that the two identities by explicitly acting the symmetry operators on the embedding matrix.

\subsection{The sewing matrices of symmetry operators}
For the unitary crystalline symmetries, \eg $g=C_{3z},C_{2x}, P, C, C'$, we define the sewing matrices as 
\begin{equation}
B_{n'n}^{(g)}(\kk) = \bra{u_{n'}(g\kk)} D(g) \ket{u_n(\kk)}.
\end{equation}
For the anti-unitary symmetries, \eg $g=C_{2z}\TRS, \PH$, we define the sewing matrices as 
\begin{equation}
B_{n'n}^{(g)}(\kk) = \bra{u_{n'}(g\kk)} D(g) \ket{u_n^*(\kk)}.
\end{equation}
Using the identities \cref{eq:VD1,eq:VD2}, we obtain that the sewing matrices are periodic in momentum space, \ie $B^{(g)}(\kk+\GG)=B^{(g)}(\kk)$ for arbitrary reciprocal lattice $\GG$. The explicit expression for the sewing matrices depend on different basis, and will be give in Ref. \cite{ourpaper3} for the cases needed for our interacting problem.

\section{More discussions on the particle-hole symmetry} \label{app:PHbreaking}

\subsection{Effect of in-plane lattice relaxation}

We first show that the in-plane relaxation does not directly lead to the particle-hole symmetry breaking. 
In Ref. \cite{Koshino2017relaxation}, Koshino et al. obtained the relaxed lattice structure of TBG by minimizing the elastic energy.
The in-plane displacement of atoms in the two layers can be approximated as 
\begin{equation} \label{eq:s-l}
\mathbf{s}^{(l)}(\rr) = \begin{cases}
\mathbf{s}(\rr),\quad & l=T\\
-\mathbf{s}(\rr),\quad & l=B
\end{cases}
\end{equation}
where $l$ is the layer index and
\begin{equation}
\mathbf{s}(\rr) =   \frac{i}2 \sum_{\GG} \mathbf{s}_\GG \, e^{i\GG\cdot\rr}\ .
\end{equation}
$\mathbf{s}_\GG$ is an odd function of $\GG$ and is dominated by the components on the shortest $\GG$ vectors \cite{Koshino2020continuum}.
For example,  for $|\GG|=|\bb_{M,1}|$, $\mathbf{s}_\GG$ takes the form \cite{Koshino2020continuum}
\begin{equation}\label{eq:s-def}
    \mathbf{s}_\GG = \frac{s_0}{|\bb_{M1}|} ( -G_y, G_x)\ ,\qquad
    ( |\GG|=|\bb_{M1}| ),
\end{equation}
where $s_0$ is the length of the displacement vectors. 
The in-plane displacement leads to two new terms in the Hamiltonian of TBG: (i) the correction of the intralayer hopping, which has a form of pseudo vector potential, (ii) the correction to the interlayer hopping.
The pseudo vector potential has the form
$\mathbf{A}$
\begin{equation}
\mathbf{A}_x(\rr) = \gamma [ s_{xx}(\rr) - s_{yy}(\rr)],\quad
\mathbf{A}_y(\rr) = -2 \gamma s_{xy}(\rr)
\end{equation}
with $s_{ij} = \frac12 (\partial_i v_j + \partial_j v_i)$ being the strain tensor and $\gamma$ the coupling constant. 
Since $\mathbf{s}_\GG$ is real and odd in $\GG$, $\mathbf{A}(\rr)$ is real and even in $\rr$. 
The pseudo vector potential enters the TBG Hamiltonian as 
\begin{equation}
 -i v_F \partial_\rr \cdot \bsigma \quad \to \quad v_F (-i\partial_\rr + A(\rr) \tau_z) \cdot \bsigma\ .
\end{equation}
The $\tau_z$ factor in the pseudo vector potential term comes from the fact  $\mathbf{s}^{(T)}=-\mathbf{s}^{(B)}=\mathbf{s}(\rr)$ (\cref{eq:s-l}). 
We find that the pseudo vector potential term respects all the symmetries. 
In particular, it respects the $C_{2z}T=\sigma_x K$ and the $\PH=i\tau_y \sigma_x K$ symmetries, which protect the stable topology.

With the in-plane displacement fields, the interlayer coupling changes to \cite{Koshino2020continuum}
\begin{align}
T(\rr) \to \tilde{T}(\rr) = \sum_{i=1}^3 e^{-i \mathbf{K}_i \cdot \mathbf{s}(\rr) - i\qq_i\cdot \rr}  \cdot  T_i \ ,
\end{align}
where $\mathbf{K}_1$ is the $\mathbf{K}$ vector shown in \cref{fig:QQ},  $\mathbf{K_2} = C_{3z} \mathbf{K_1}$, $\mathbf{K_3}=C_{3z} \mathbf{K_2}$, and $T_i$ are given in \cref{eq:Ti-def}. 
The corresponding Hamiltonian matrix can be written as 
\begin{equation}
\begin{pmatrix}
0 & \tilde{T}(\rr) \\
\tilde{T}^\dagger (\rr) & 0
\end{pmatrix} = \sum_i \pare{ f_i(\rr) \tau_x \otimes T_i + g_i(\rr) \tau_y \otimes T_i }\ ,
\end{equation}
with $f_i(\rr)$ and $g_i(\rr)$ being the real part and imaginary part of $e^{-i \mathbf{K}_i \cdot \mathbf{s}(\rr) - i\qq_i\cdot \rr} $, respectively. 
Since $s(\rr)$ is odd in $\rr$, there is $f_i(\rr) = f_i(-\rr)$ and $g_i(\rr) = -g_i(\rr)$. 
One can find that the interlayer coupling respects both $C_{2z}T=\sigma_x K$ and $\PH=i\tau_y \sigma_x K$. 

Therefore, the in-plane relaxation does not directly lead to particle-hole symmetry breaking. 
This is also numerically shown in Refs. \cite{Koshino2020continuum} and \cite{fang2019angle}.

\subsection{\texorpdfstring{$\kk$}{k}-dependence in the interlayer hopping}

The continuous models of TBG discussed in the rest of this work have made an approximation on the interlayer coupling: We have neglected the $\kk$-dependence of the Fourier transformation of the interlayer hopping. 
Refs. \cite{Koshino2020continuum} and \cite{fang2019angle} have shown that keeping the $\kk$-dependence in $t(\qq)$ will lead to a particle-hole asymmetry in the energy spectrum. 
Ref. \cite{fang2019angle} shows that keeping linear order of $\kk$ in the $\kk$-dependence is a good approximation.
Here we derive the linear $\kk$-dependence explicitly. 
Following Eq. (14) of the supplementary material of Ref. \cite{song_all_2019}, we have the intervalley coupling on the basis \cref{eq:Hk-basis1,eq:Hk-basis2}
\begin{equation}\small
H_{\QQ,\QQ'}(\kk) = \sum_{j=1}^3 t_{\mathbf{K}_j+\kk-\QQ} \delta_{\QQ,\QQ'-\qq_j} \cdot \bar{T}_j\quad (\QQ\in\mathcal{Q}_T,\QQ'\in\mathcal{Q}_B)
\end{equation}
where $t_\qq = t_0 \exp( - \alpha (|\qq|d_\perp)^\gamma )$ is the Fourier transformation of the interlayer coupling with $d_\perp$ being the distance between the two layers, and 
\begin{equation}
\bar{T}_i = \sigma_0 + \Big[\sigma_x\cos\frac{2\pi(i-1)}3+ \sigma_y\sin\frac{2\pi(i-1)}3\Big] \ .
\end{equation}
Here $\mathbf{K}_1$ is the $\mathbf{K}$ vector shown in \cref{fig:QQ},  $\mathbf{K}_2 = C_{3z} \mathbf{K}_1$, $\mathbf{K}_3=C_{3z} \mathbf{K}_2$.
The parameters in $t_\qq$ have been fitted as $\alpha=0.13$, $\gamma=1.25$, , 
For simplicity here we assume $d_\perp$ is uniform in real space. 
\begin{figure}
    \centering
    \includegraphics[width=0.9\linewidth]{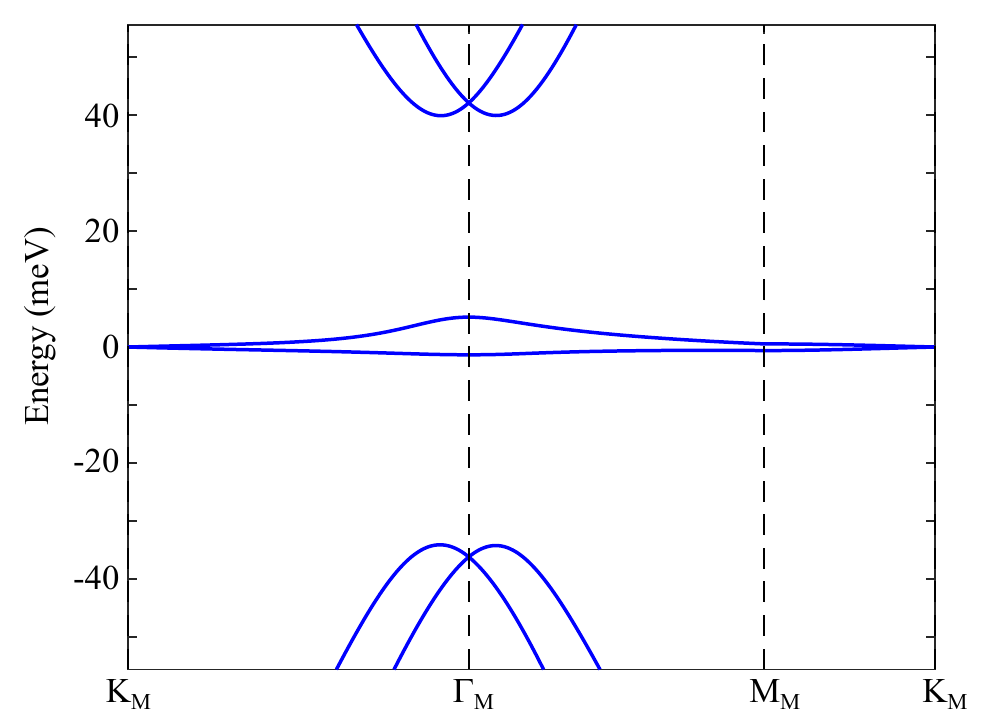}
    \caption{Significant particle-hole asymmetry in the energy spectrum when the $\kk$-dependence of the interlayer coupling is considered. The parameters of Hamiltonian used in (d-g) are $v_F=5.944{\rm eV \cdot \mathring{ A} }$, $|K|=1.703\mathring{\rm A}^{-1}$, $w_1=110{\rm meV}$, $w_0=0.7w_1$, $\lambda=250 {\rm meV}$, $\theta=1.05^\circ$. }
    \label{fig:band-PHA}
\end{figure}

We expand $t_{\mathbf{K}_j+\kk-\QQ}$ to leading order of $\kk-\QQ$ as 
\begin{equation}
t_{\mathbf{K}_j+\kk-\QQ} \approx t_{\mathbf{K}_j} \pare{ 1 - \alpha \gamma (d_\perp|\mathbf{K}_j|)^\gamma \frac{\mathbf{K}_j\cdot (\kk-\QQ)}{|\mathbf{K}_j|^2} }
\end{equation}
Then we obtain the correction to the continuous model as 
\begin{equation}\small
\delta H_{\QQ,\QQ'}(\kk) = - \lambda \sum_{j=1}^3 \delta_{\QQ,\QQ'-\qq_j}\cdot \frac{\mathbf{K}_j\cdot (\kk-\QQ)}{|\mathbf{K}_j|^2} \cdot \bar{T}_j
\end{equation}
for $\QQ\in\mathcal{Q}_T,\QQ'\in\mathcal{Q}_B$, and 
\begin{equation}\small
\delta H_{\QQ,\QQ'}(\kk) = - \lambda \sum_{j=1}^3 \delta_{\QQ,\QQ'+\qq_j}\cdot \frac{\mathbf{K}_j\cdot (\kk-\QQ')}{|\mathbf{K}_j|^2} \cdot \bar{T}_j
\end{equation}
for $\QQ\in\mathcal{Q}_B,\QQ'\in\mathcal{Q}_T$.
Here $\lambda= t_{\mathbf{K}} \alpha \gamma (d_\perp|\mathbf{K}|)^\gamma$. 
Since $\delta H_{\QQ,\QQ'}(\kk)$ still consists of $\sigma_0$, $\sigma_x$, $\sigma_y$, it respects the $C_{2z}T$ symmetry. 
However, it breaks the unitary particle-hole symmetry $P$ (\cref{eq:P-action}) and hence the anti-unitary particle-hole symmetry $\PH= P\cdot C_{2z}T$. 
We use the parameter $\lambda= 180 \mathrm{meV}$ to make a significant particle-hole asymmetry in the band structure: the lower band at $\Gamma_M$ has the energy $-1.34$meV and the upper band has the energy 5.16meV (\cref{fig:band-PHA}).
However, we find that the error of the particle-hole symmetry for the Bloch wavefunctions (\cref{eq:errorPH}) as $\mathrm{error}(\PH) = 0.07$, which is still weak. 
A more realistic model should include both the in-plane relaxation and the $\kk$-dependence of the interlayer hopping \cite{Koshino2020continuum,fang2019angle}.
As discussed in the last subsection, without the $\kk$-dependence, the in-plane relaxation will not lead to $\PH$-breaking. 
Hence we claim that the error of the $\PH$ symmetry in this more realistic model will have the same order as $0.13$.

% \begin{align}
% \delta T(\rr) \approx & 
% \frac{i}2\begin{pmatrix}
% \lambda_1 \{ t_-(\rr), \hat{k}_+ \} & 0 \\
% \lambda_2 \{ t_+(\rr), \hat{k}_+ \} & \lambda_1 \{ t_-(\rr), \hat{k}_+ \} 
% \end{pmatrix} \nono\\
% -& \frac{i}2  \begin{pmatrix}
% \lambda_1 \{ t_+(\rr), \hat{k}_- \} & \lambda_2 \{ t_-(\rr), \hat{k}_- \} \\
% 0 & \lambda_1 \{ t_+(\rr), \hat{k}_- \}
% \end{pmatrix}\ ,
% \end{align}
% where $\lambda_2=2\lambda_1=0.18 \mathrm{eV\cdot\mathring{A}}$, $\hat{k}_\pm = -i \pt_x \pm \pt_y$ and 
% \begin{equation}
%     t_0(\rr) = e^{-i\qq_1\cdot\rr} + e^{-i\qq_2\cdot\rr} + e^{-i\qq_3\cdot\rr}\ ,
% \end{equation}
% \begin{equation}
%     t_-(\rr) = e^{-i\qq_1\cdot\rr} + e^{i\frac{2\pi}3-i\qq_2\cdot\rr} + e^{-i\frac{2\pi}3-i\qq_3\cdot\rr}\ ,
% \end{equation}
% \begin{equation}
%     t_+(\rr) = e^{-i\qq_1\cdot\rr} + e^{-i\frac{2\pi}3-i\qq_2\cdot\rr} + e^{i\frac{2\pi}3-i\qq_3\cdot\rr}\ .
% \end{equation}

\section{The gauge of Chern band basis}\label{app:Chern-gauge}

In \cref{sec:Chern} we have explained that the choice of Chern band basis due to the $C_{2z}\TRS$ gauge fixing (\cref{eq:Chern-basis}) has two ambiguities: (i) the choice of the two branches of $\frac{\theta_n(\kk)}{2}$, \ie $\frac{\theta_n(\kk)}{2}$ and $\frac{\theta_n(\kk)}{2}+\pi$, and (ii) the choice of $\ket{u_{1}(\kk)}$ and $\ket{u_{-1}(\kk)}$ at Dirac points where the two bands are degenerate. 
Both ambiguities lead to an ambiguity when choosing $\ket{v_\pm(\kk)}$.
If we replace $\frac{\theta_{1}(\kk)}2$ by $\frac{\theta_{1}(\kk)}2+\pi$ in \cref{eq:Chern-basis}, then the two Chern band states in the new gauge $\ket{v_\pm'(\kk)}$ are related to the Chern band states in the previous gauge as
\begin{align}
\ket{v_\pm'(\kk)}=&\frac{1}{\sqrt2}( - e^{i\frac{\theta_1(\kk)}2} \ket{u_1(\kk)} \pm i e^{i\frac{\theta_{-1}(\kk)}2} \ket{u_{-1}(\kk)})\nono\\
=& - \ket{v_\mp(\kk)}.
\end{align}
Therefore, changing the branch of $\frac{\theta_1(\kk)}{2}$ at $\kk$ will interchange the two states $\ket{v_\pm(\kk)}$ at $\kk$.
One can show that changing the branch of $\frac{\theta_{-1}(\kk)}{2}$ will also interchange the two states $\ket{v_\pm(\kk)}$ for the same reason.
We then consider to interchange  $\ket{u_{\pm1}(\kk)}$ at a Dirac point $\kk_D$.
The Chern band states $\ket{v_\pm''(\kk_D)}$ defined with the interchanged $\ket{u_{\pm1}(\kk_D)}$ become
\begin{align}
\ket{v_\pm''(\kk_D)}=&\frac{1}{\sqrt2}( e^{i\frac{\theta_{-1}(\kk_D)}2} \ket{u_{-1}(\kk_D)} \pm i e^{i\frac{\theta_{1}(\kk_D)}2} \ket{u_{1}(\kk_D)})\nono\\
=& \pm i \ket{v_\mp(\kk)}.
\end{align}
Thus interchanging $\ket{u_{\pm1}(\kk)}$ at Dirac points will interchange $\ket{v_\pm(\kk)}$ at the Dirac points. 

Due to the ambiguities discussed above, at each $\kk$ point we only have two choices for $\ket{v_{\pm}(\kk)}$ (up to phase factors). 
Starting with $\ket{v_{\pm}(\kk_0)}$ at a given momentum $\kk_0$, the condition \cref{eq:Chern-band-gauge} will uniquely determine the two branches of $\ket{v_{\pm}(\kk)}$ over the whole BZ. 
To be specific, we consider the choice of $\ket{v_{\pm}(\kk_0+\qq)}$ for $\qq$ being a small momentum. 
Suppose $\ket{v_{\pm}'(\kk_0+\qq)}$ is returned by \cref{eq:Chern-basis} without considering \cref{eq:Chern-band-gauge}.
Then if they satisfy $|\inn{v_{m}(\kk_0)| v'_{m'}(\kk_0+\qq)}|\approx \delta_{mm'}$, we choose $\ket{v_{\pm}(\kk_0+\qq)}=\ket{v_{\pm}'(\kk_0+\qq)}$; otherwise, \ie $|\inn{v_{m}(\kk_0)| v'_{m'}(\kk_0+\qq)}|\approx 1-\delta_{mm'}$, we choose $\ket{v_{\pm}(\kk_0+\qq)}=\ket{v_{\mp}'(\kk_0+\qq)}$.
Repeating this procedure over the whole BZ iteratively will uniquely determined the two branches of $\ket{v_{\pm}(\kk)}$.

We now explain that the obtained two branches by the method introduced above must be Chern bands.
We first consider $\kk$ not at the Dirac points. 
Due to the discussion in \cref{sec:Chern}, if a smooth branch of $\frac{\theta_n(\kk)}{2}$ is chosen, the states $\ket{v_{\pm}(\kk)}$  will have smooth Berry's curvatures $\pm f(\kk)$.
On the other hand, if the branch of $\frac{\theta_n(\kk)}{2}$ changes at $\kk_\star$, then there is $\lim_{\qq\to0} |\inn{v_{+}(\kk_\star)|v_+(\kk_\star+\qq)}|=|\inn{v_{+}(\kk_\star)|v_-(\kk_\star)}|=0$ and hence the Berry's curvatures of $\ket{v_\pm(\kk)}$ will be discontinuous at $\kk_\star$. 
Therefore, choosing a smooth branch of $\frac{\theta_n(\kk)}{2}$ is equivalent to choosing $\ket{v_\pm(\kk)}$ such that they have continuous Berry's curvatures.
Thus \cref{eq:Chern-band-gauge}, which guarantees continuous Berry's curvatures, also enforces that $\ket{v_\pm(\kk)}$ must have the Berry's curvatures $\pm f(\kk)$.
We then consider the Dirac points. 
Since there is a finite number of Dirac points in the BZ, the integral of Berry's curvature in the (infinite small) neighborhoods of the Dirac points approaches zero as long as \cref{eq:Chern-band-gauge} holds such that the Berry's curvature is non-divergent. 
Therefore, the integral of Berry's curvatures of $\ket{v_\pm(\kk)}$ subject to the condition \cref{eq:Chern-band-gauge} are $\pm \frac1{2\pi}\int d^2\kk f(\kk)=\pm e_2$.

Now we explicitly show why the Berry's curvatures of \cref{eq:Chern-basis} are not divergent at the Dirac points if \cref{eq:Chern-band-gauge} holds.
We consider a linearized k$\cdot$p model around a single Dirac point $H = k_x \sigma_x + k_y \sigma_y$, where the $C_{2z}\TRS$ operator is $\sigma_x K$.
Since Berry's curvature is gauge-invariant, we take the particular gauge $\theta_{1}(\kk)=\theta_{-1}(\kk)=0$ in the calculation.
The Bloch states in this gauge are
{\small
\begin{equation}
\ket{u_{1}(\kk)} = \frac1{\sqrt2}\begin{pmatrix}
	e^{-i\frac{\phi(\kk)}2}\\ e^{i\frac{\phi(\kk)}2}
	\end{pmatrix},\quad
\ket{u_{-1}(\kk)} = \frac1{\sqrt2}\begin{pmatrix}
	-i e^{-i\frac{\phi(\kk)}2}\\ i e^{i\frac{\phi(\kk)}2}
	\end{pmatrix},
\end{equation}
}%
where $\phi(\kk)=\arccos\frac{k_x}{\sqrt{k_x^2+k_y^2}}$.
Then the recombined Chern basis are $\ket{v_{+}(\kk)} =(e^{-i\frac{\phi(\kk)}2}, 0)^T$, $\ket{v_{-}(\kk)}=(0,e^{i\frac{\phi(\kk)}2})^T$.
They satisfy \cref{eq:Chern-band-gauge} and have vanishing Berry's curvatures in the neighbourhood of the Dirac point.

\clearpage

\end{document}